\documentclass[a4paper,11pt]{article}

\usepackage{graphicx}
\usepackage{amssymb}
\usepackage{amsmath}
\begin{document}

\title{Atmospheric multiple scattering of fluorescence and Cherenkov light emitted by extensive air showers}
\author{J. P\c ekala, P. Homola, B. Wilczy\'nska, H. Wilczy\'nski\\
Institute of Nuclear Physics PAN, Radzikowskiego 152, 31-342 Krak\'ow, Poland}
\maketitle

\begin{abstract}
Atmospheric scattering of light emitted by an air shower not only attenuates direct fluorescence light from the shower, but also contributes to the observed shower light. So far only direct and singly-scattered Cherenkov photons have been taken into account in routine analyses of the observed optical image of air showers. In this paper a Monte Carlo method of evaluating the contribution of multiply scattered light to the optical air shower image is presented, as well as results of simulations and a parameterization of scattered light contribution to measured shower signal.
\end{abstract}


\section{Introduction}

Charged particles of an extensive air shower produce a large number of fluorescence and Cherenkov photons on their way through the atmosphere. Fluorescence light coming directly from the shower to the optical detector provides information needed for determining the number of particles in the shower at points along its trajectory. The profile of longitudinal shower development (i.e. the number of particles in the shower versus depth in the atmosphere) is thus obtained. This is the principle of the fluorescence method of air shower detection. Photons emitted by the shower may get scattered on their way through the atmosphere between the shower front and the detector. Single scattering deflects the photons and therefore results in attenuation of the photon beam arriving to the detector. This effect is routinely accounted for in experimental air shower studies. However, photons may also undergo a {\em series} of scatterings before reaching finally the optical detector so that some of them may get redirected again toward the detector. Thus the atmospheric scattering results not only in attenuation of the direct fluorescence signal, but may also contribute to the signal received by the detector. Since the path of multiply-scattered photons is longer than the direct shower-detector distance, these photons will be delayed with respect to direct fluorescence photons which arrive without scattering. The scattered light (both fluorescence and Cherenkov) must be regarded as a background for the direct fluorescence signal, because its intensity relates to the history of shower development rather than to the current number of particles in the shower.

In most cases, the optical image of a shower recorded by detectors consists mainly of direct fluorescence and singly-scattered Cherenkov photons. Cherenkov photons are emitted mostly at small angles with respect to the shower axis, their paths are very close to the shower itself, and they accumulate as the shower develops. Thus singly scattered Cherenkov photons can make a large contribution to the shower image, as they arrive the detector simultaneously and from approximately the same direction as fluorescence photons, making discerning them difficult. A correction for the singly scattered Cherenkov photons is a standard part of all procedures of shower reconstruction based on fluorescence observations. Direct Cherenkov light makes a significant contribution to the signal in the fluorescence detector only when the shower lands close to the detector site or for very inclined showers \cite{baltrusaitis,plamka}. Scattered light may be relatively strong in all geometrical configurations, especially in the late stages of shower development.

Fluorescence photons, produced by the shower may also undergo scattering. A small part of the singly scattered photons get to the detector and are recorded together with the direct fluorescence light. Remaining photons, after traveling some distance in the atmosphere may scatter again, this time in direction of the detector. Though not likely, it is possible that a fraction of shower photons may reach the detector after a longer series of scatterings \cite{tsukuba}.

Only recently first attempts have been made to account for scattered fluorescence photons or multiple scattering of Cherenkov light in shower reconstruction procedures \cite{gap2008052}. Few studies of this effect have been done so far. In \cite{roberts} multiple scattering of fluorescence photons was investigated for vertical showers and an estimate of the contribution of the scattering effect to the shower image was obtained. An analytical study of Rayleigh scattering of fluorescence photons was made in \cite{giller}. A comprehensive study of scattering of light in the atmosphere is so far missing.

A good estimate of the contribution of the scattering effect to the shower optical image is needed for precise shower reconstruction. The multiple scattering effect increases the amount of light arriving from the shower to the detector. Failure to account for this effect results in a systematic error in shower energy determination in the fluorescence method of detection.

The experimental energy spectra determined by different experimental techniques (surface array and fluorescence detector) do not coincide \cite{yoshida}. The differences most probably are due to systematic uncertainties in shower reconstruction in both techniques. Systematic errors in the fluorescence technique are caused mainly by uncertainties in fluorescence yield, detector calibration, shower reconstruction and atmospheric effects - which include the effect of the multiple scattering. Reduction of these systematic errors is of primary importance to the studies of cosmic ray energy spectrum.

The objective of this paper is the systematic study of the multiple scattering effect, by both Rayleigh and Mie scattering processes. A correction to the existing shower reconstruction procedures is obtained to account for the scattering effect.

\section{Method of simulation}

Multiple scattering of fluorescence and Cherenkov photons was simulated using the ``Hybrid\_fadc'' program \cite{dawson}, which was designed to simulate air shower development and detection. In this program, calculations are done in steps corresponding to a change of 0.04$^{\circ}$ in shower position on the sky, as seen by the detector. For chosen primary particle energy and shower geometry, in each step the program calculates the shower size using the Gaisser-Hillas parameterization and the number of emitted fluorescence and Cherenkov photons. Based on these, the number of Cherenkov photons (both emitted directly towards the detector and singly-scattered towards it) are calculated. The number of fluorescence photons emitted towards the detector is also determined. The shower is assumed to have no lateral distribution, i. e. all photons are emitted at the shower axis. Calculations are done in 16 wavelength bins covering the range from 276 nm to 420 nm. Initially, the default ``Hybrid\_fadc'' settings were used for angular distribution of Cherenkov emission, molecular atmosphere profile, aerosol distribution and detector location. Next alternatives to these settings were tested in dedicated sets of simulations - their impact on the results is discussed later.

\begin{figure}[tp]
\begin{center}
\includegraphics[scale=0.5]{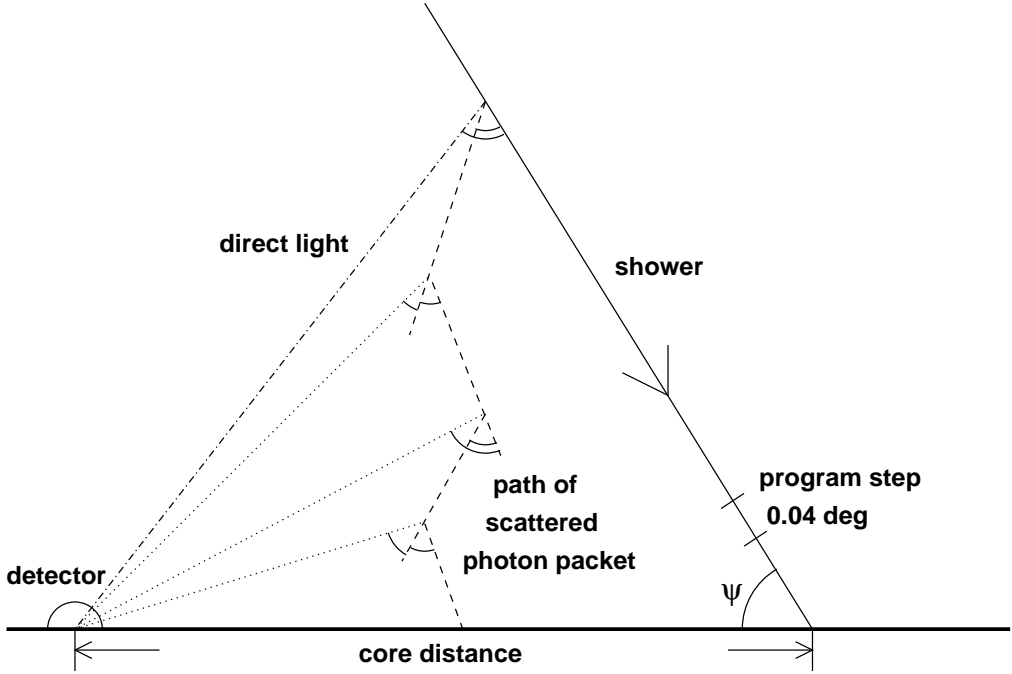}
\caption{\label {fig0} Method of simulation (see text for details).}
\end{center}
\end{figure}

In order to simulate multiple scattering of fluorescence and Cherenkov photons, modifications to the original program were made. Due to large number of photons involved, in calculations concerning photons produced in a given step it is impossible to trace them all separately, so that some simplifications are inevitable. The total number of photons is divided into smaller ``packets'' (typically 10 thousand packets in each step and each wavelength bin). All following calculations are done for each packet.

A packet starts from the shower axis at the point corresponding to a current shower development step, with its direction randomly chosen, based on either an isotropic angular distribution (for fluorescence photons) or an exponential one ($dN/d\Omega\propto e^{-\theta/\theta_{0}}/2\pi sin\theta$, with $\theta_{0} \simeq 4.5^{\circ}$) for Cherenkov light \cite{baltrusaitis}. Assuming that all photons in a packet scatter at one point, two distances to scattering points, for Rayleigh and Mie scattering processes, are randomly drawn from the corresponding mean free path distributions. From the two calculated points, the one closer to the starting point is chosen as the place where the scattering occurs. Knowing the geometry of the event and also the angular distribution of scattered light ($d\sigma/d\Omega\propto1+cos^{2}\theta$ for Rayleigh or approximately $d\sigma/d\Omega\propto e^{-\theta/26.7^{\circ}}$ for Mie scattering \footnote{There is a large-angle tail of this distribution. Details of its parameterization were shown in \cite{roberts} to have little effect on the overall scattering effect.}, respectively), the number of photons scattered towards the detector is found. Next, including the attenuation factor for the path toward the eye, the signal at the detector due to the scattered portion of the packet is calculated. With the information about the whole path in the atmosphere, the time of arrival of the scattered photons to the detector is found. In order to trace the remaining photons in the packet, it is assumed that they continue their flight together. For this smaller packet, a direction is randomly chosen and all calculations, just as for the first scattering, can be repeated several times. These calculations give as output the information about each packet: size of signal, arrival direction on the sky and time of arrival to the detector.

It may happen that the point of scattering is placed by chance very close to the detector. In such a case, the large solid angle of the detector, as seen from this point causes the signal from this single scattering to be relatively very strong. In simulation results these cases can be recognized by the values of signal much larger than from typical packet fractions arriving from neighboring directions. These are artefacts of the simulation method and can be recognized in the results shown below.

To investigate the instantaneous image of the shower, the signal from photons arriving simultaneously, i. e. within one program step (corresponding to a change of 0.04$^{\circ}$ in shower position on the sky) is integrated. The duration of the program step (integration time) varies from about 2 ns to about 40 ns, depending on shower geometry. These are short time intervals, compared to the signal integration time in a real detector. Moreover, in this paper we present ratios of the scattered to direct signal, so the variation of the signal integration time does not influence the results. For all time intervals of the program steps, the signal from direct and scattered light is calculated separately.

\section{Simulation results}

In the following, the term \textit{``new signal''} on all plots denotes scattered (both singly and multiply) fluorescence plus multiply scattered Cherenkov light; the \textit{``shower''} or \textit{``old signal''} denotes direct fluorescence plus direct and singly scattered Cherenkov photons; the \textit{``total signal''} - the sum of these signals.

\begin{figure}[tp]
\begin{center}
\includegraphics[scale=0.9]{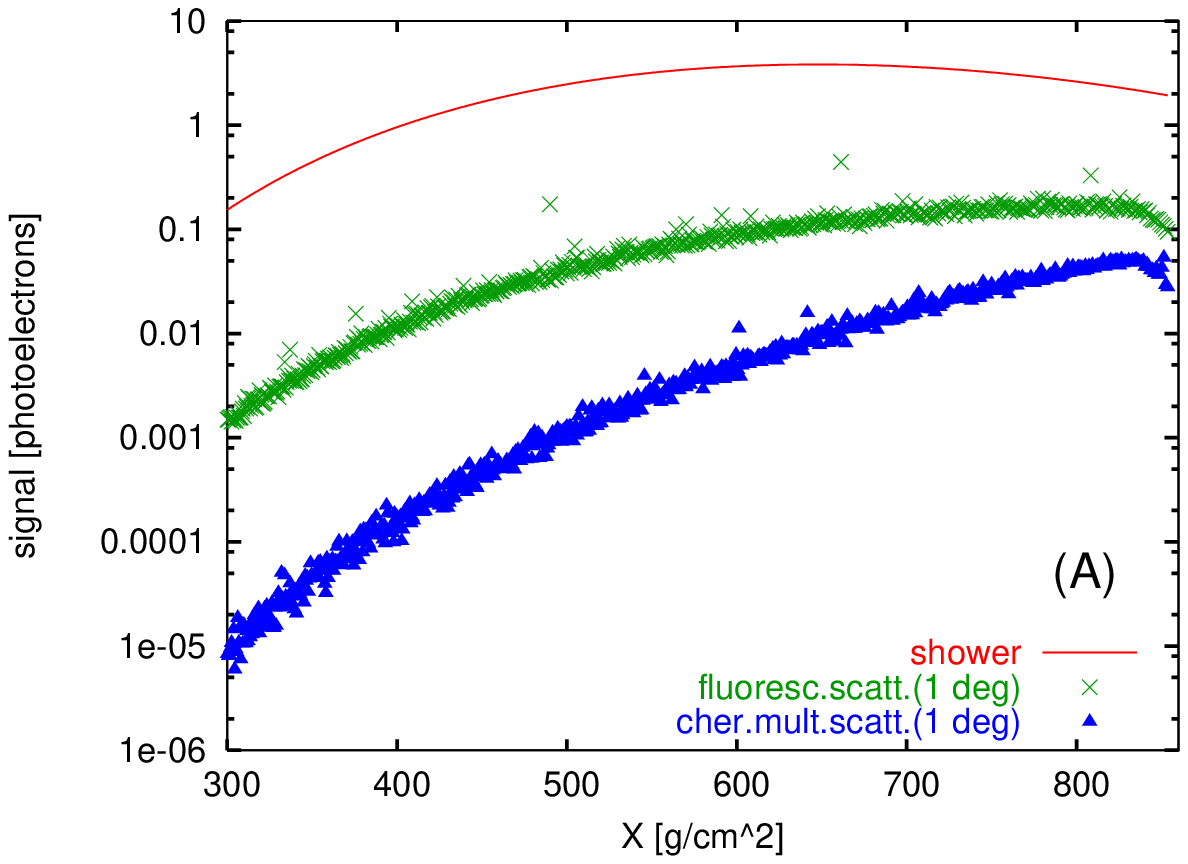}\\
\includegraphics[scale=0.9]{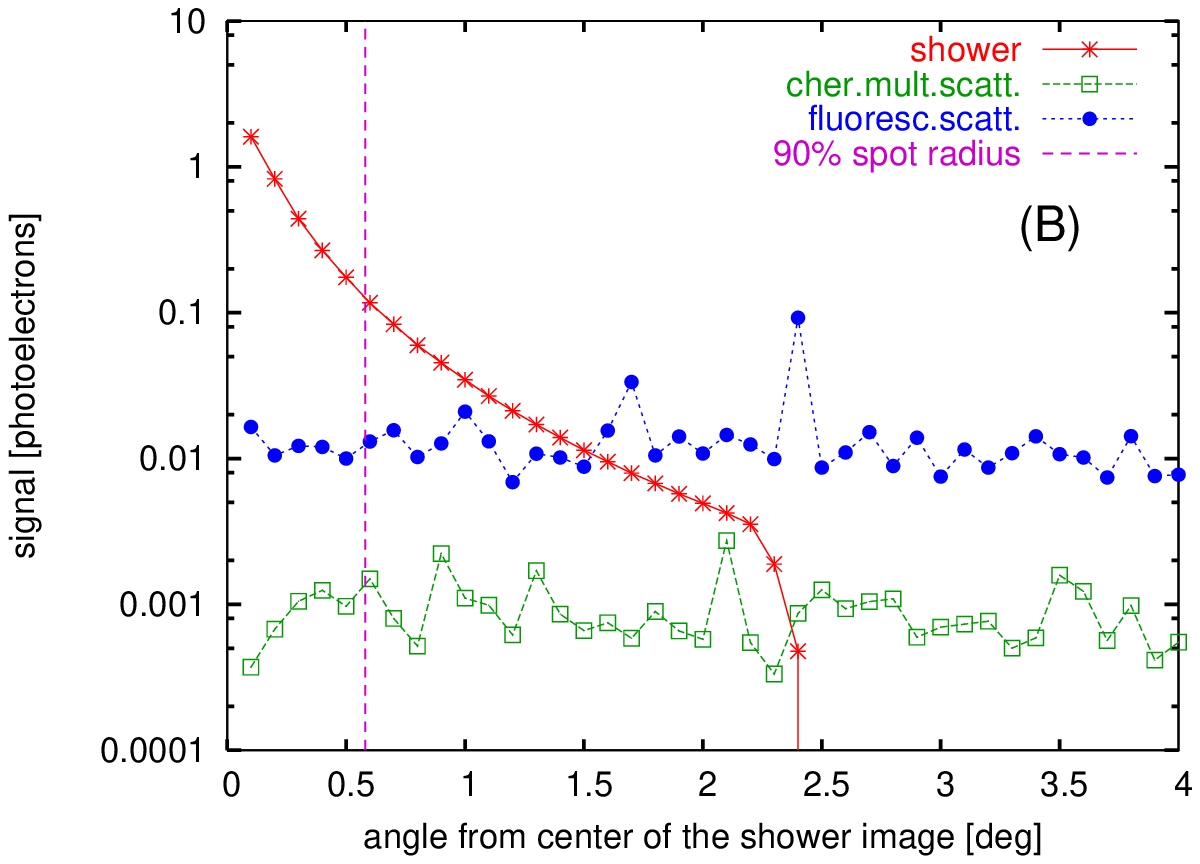}\\
\linespread {1.}
\caption{\label {fig2} Example of simulation results (vertical $10^{19}$eV shower landing 15 km from detector). (A) Shower longitudinal profile and scattered light contribution in 
1$^{\circ}$-radius circle around momentary shower position is shown. (B) The distribution of light on the sky (integrated in rings 0.1$^{\circ}$ wide) in shower maximum is shown. The vertical line marks the radius of the spot containing 90\% of the signal.}
\end{center}
\end{figure}

With the program described above, simulations were made for different shower configurations, namely for all combinations of:\\
$\cdot$ energy - $10^{18}$, $10^{19}$, $10^{20}$, $10^{21}$eV;\\
$\cdot$ core distance - 3, 7, 15, 25 km;\\
$\cdot$ ${\psi}$ angle (see Fig. \ref{fig0}) within the shower-detector plane (SDP) - 30, 50, 70, 90, 110, 130, 150 degrees for vertical SDP;\\
$\cdot$ SDP inclination - 30, 45, 60, 70 degrees with ${\psi}=90^{\circ}$.\\
This set of simulations was done first using the US Standard Atmosphere Model \cite{usstd} and one value of aerosol concentration, corresponding to total horizontal attenuation length $\Lambda_{T}$=6.347 km. It corresponds to a high aerosol concentration in the air - this allows us to investigate the effect of the multiple scattering in conditions, in which it is most prominent. Simulations with other aerosol concentrations were also done, and their results are presented below.

\begin{figure}[tp]
\begin{center}
\includegraphics[scale=0.95]{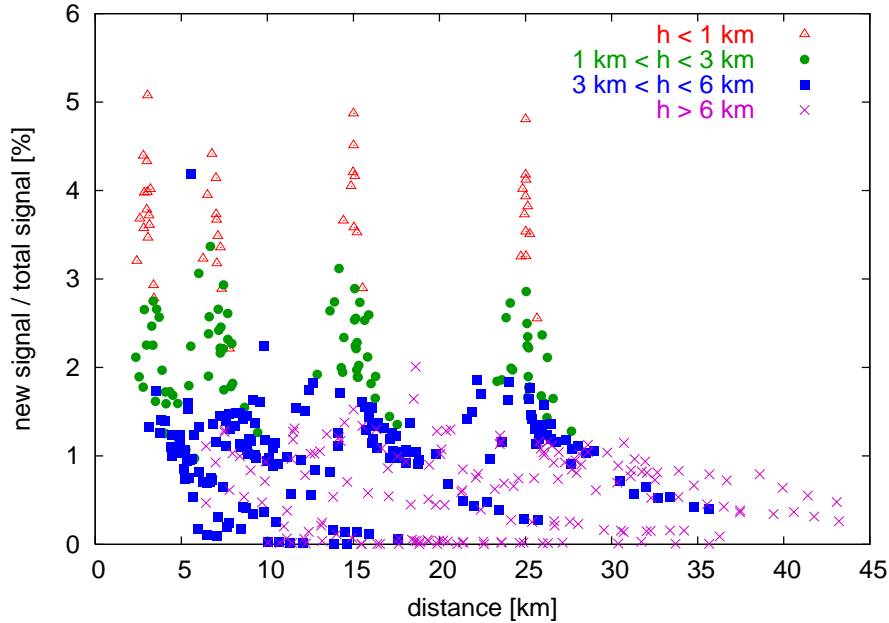}
\caption{\label {fig3} Contribution of ``new signal'' to total shower signal versus the shower-detector distance, in different altitude ranges (within the image spot containing 90\% of signal). The ``new signal'' includes scattered fluorescence and multiply scattered Cherenkov photons. The points at low altitudes have largest contributions of the scattered light, and are grouped at distances corresponding to chosen core distances of simulated showers.}
\end{center}
\end{figure}

An example of results from a single simulation run is shown in Fig. \ref{fig2}. Shown are contributions to light arriving to a ground detector: the intensity of light along the observed shower track versus its depth in the atmosphere (shower longitudinal profile). The ``shower'' curve includes direct fluorescence, direct and singly scattered Cherenkov photons. In addition the contribution of scattered (including single scattering) fluorescence photons is shown, as well as the contribution from multiple scattering of Cherenkov photons. The contribution of scattered fluorescence light (including single scattering) is at all stages larger than from multiple scattering of Cherenkov light. The signal from multiply scattered light is larger at later stages of shower development, and may finally reach a few percent of the total signal from the shower.
To compare contributions of scattered light in various shower geometries, in Fig. \ref{fig3} shown are contributions of scattered fluorescence and multiply scattered Cherenkov light within the ``image spot'', i. e. a circle containing 90\% of light from the air shower image \cite{plamka}.

\begin{figure}[tp]
\begin{center}
\includegraphics[scale=0.95]{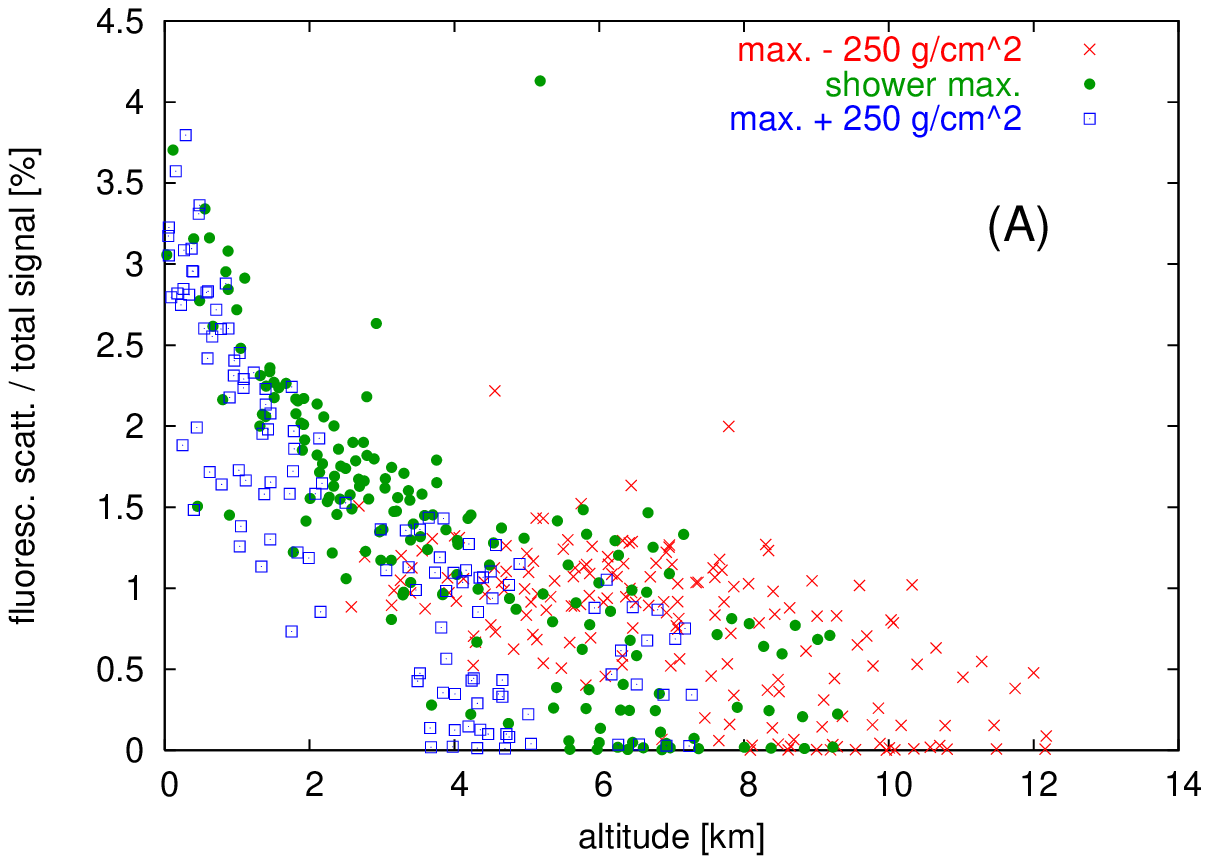}\\
\includegraphics[scale=0.95]{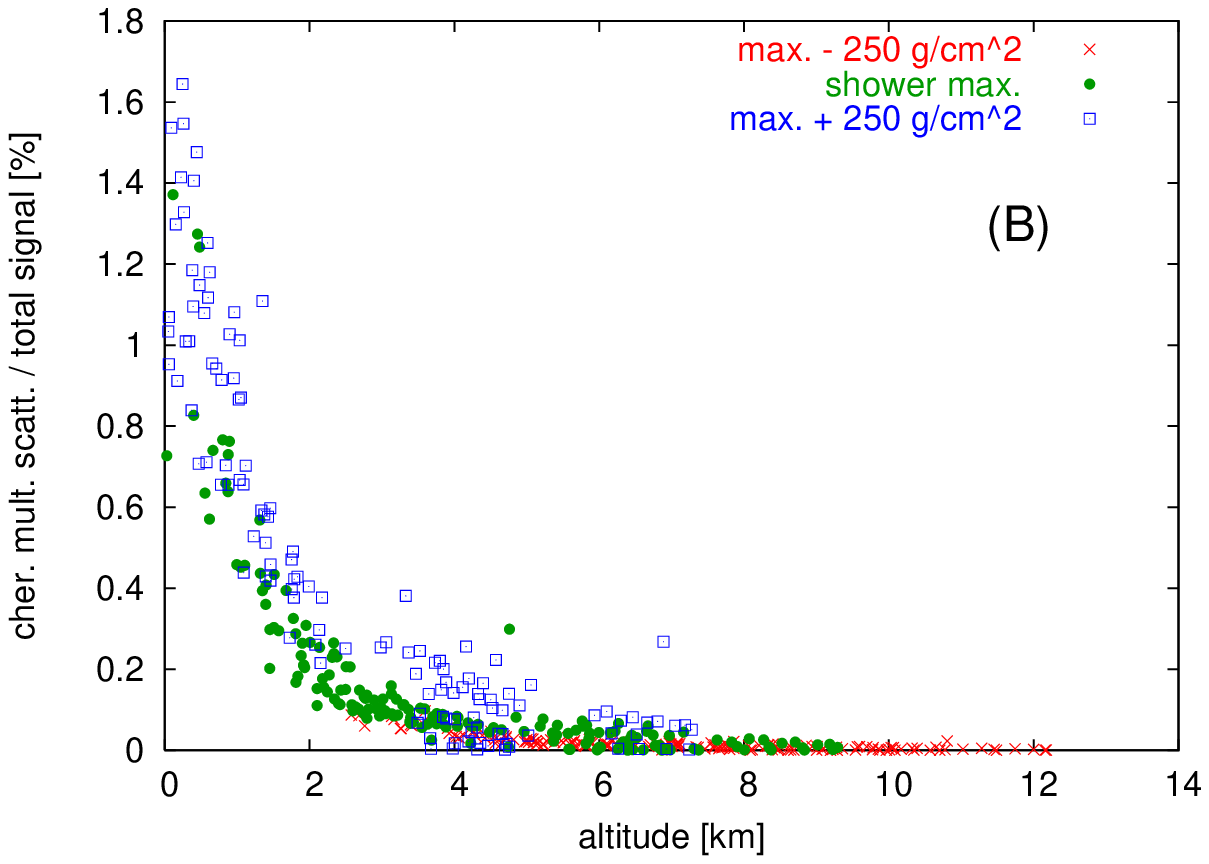}\\
\caption{\label {fig4} Contribution of singly and multiply scattered fluorescence (A) and multiply scattered Cherenkov (B) to total shower signal plotted versus altitude above ground (within the image spot containing 90\% of signal).}
\end{center}
\end{figure}

Results from the whole set of simulations performed are presented in Figures \ref{fig3} and \ref{fig4}. Shown are contributions from multiply scattered light for all showers at their maxima and points 250 g/cm$^{2}$ before and 250 g/cm$^{2}$ after the maximum (if above ground). In Fig. \ref{fig3} the simulation points group around 3, 7, 15 and 25 km distance because of chosen core distances in simulations. The relative contribution of the scattered light appears to be strongly correlated with altitude above ground, rather than with distance from shower to detector. This contribution increases with decreasing altitude. Figure \ref{fig3} suggests therefore that the altitude above ground rather than the shower-detector distance may be a better parameter to organize the data. The same results are plotted in Fig. \ref{fig4} as a function of altitude. The contribution of scattered fluorescence photons and multiply scattered Cherenkov photons are shown separately. The contributions from shower maxima and points earlier and later in shower development that are at the same altitude, show also comparable contribution of scattered light - no clear dependence on shower age is seen. On the plots shown are relative contributions of multiply scattered light to the shower image, within the 90\% ``image spot''. When calculated in this form, the scattering contribution does not depend on the size of the shower image nor on its brightness. The stochastic nature of the scattering process, enhanced by the packet algorithm and limitation of available simulation time, lead to a relatively large spread of points.

\begin{figure}[tp]
\begin{center}
\includegraphics[scale=0.95]{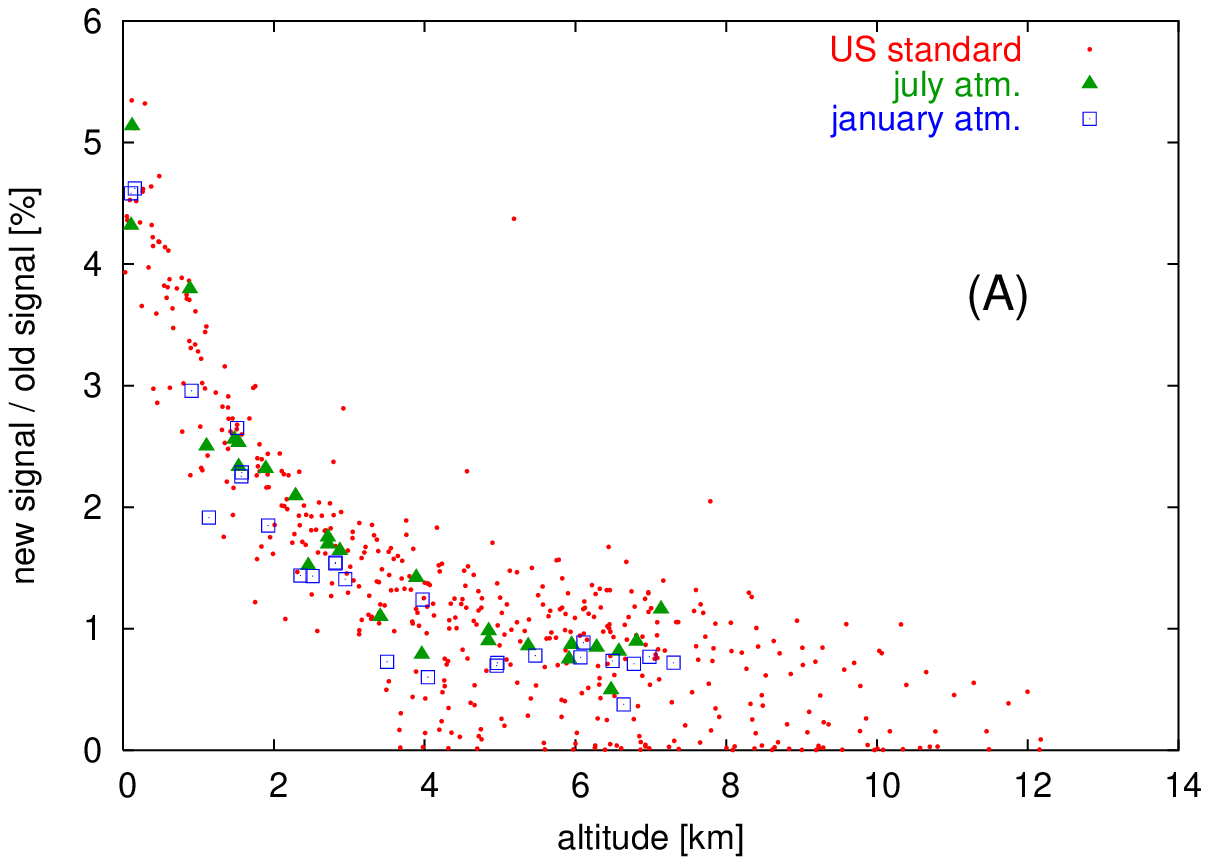}\\
\includegraphics[scale=0.95]{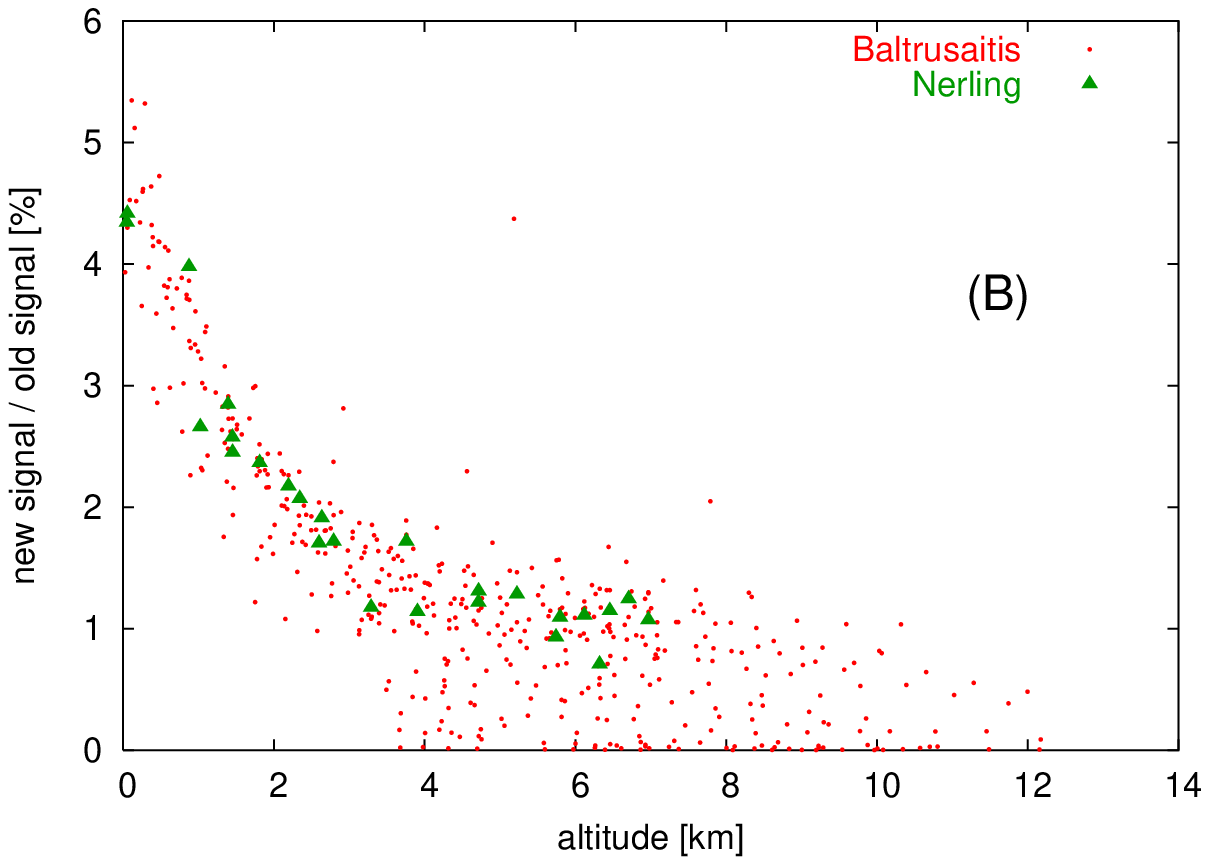}\\
\caption{\label {fig5} Comparison of simulation results for different atmospheric models (A) and different angular distributions of Cherenkov emission (B).}
\end{center}
\end{figure}

To investigate the impact of the vertical distribution of the atmosphere (the so-called molecular atmospheric profile) on the contribution of multiply scattered light, smaller sets of simulations were made. Compared were the US Standard Atmosphere Model (used in all previous simulations) and the models of atmosphere in January and July at the southern Pierre Auger Observatory in Malarg\"ue (Argentina) \cite{atmosf,atmosf2}. Results of the simulations show that changing the molecular atmospheric models has little influence on the final results, as shown in Fig. \ref{fig5}A: the effect of changing the atmospheric profile is smaller than the scatter of points due to variation of other parameters. In other words, the dependence on the atmospheric profile is a second order effect.

Compared were also results of simulations using different angular distributions of Cherenkov photons emitted by a shower: simple, one-exponential distribution: $dN/d \Omega \propto e^{-\theta/4.5^{\circ}}/2 \pi sin \theta$ \cite{baltrusaitis} with a more realistic two-exponential one \cite{nerling}:
 
$$ dN/d\Omega \propto e^{-\theta/4.5^{\circ}}/2 \pi sin \theta, \theta<35^{\circ}\mbox{ and }
 dN/d \Omega \propto e^{-\theta/22.5^{\circ}}/2 \pi sin \theta, \theta>35^{\circ} $$

The results are shown in Fig. \ref{fig5}B. Again, no significant difference between these two simulation sets can be seen. Both models of Cherenkov emission differ only in distribution of a very small fraction of photons emitted at large angles. On the other hand, the distribution of photons after a series of scatterings is not expected to be very sensitive to fine details of the original angular distribution of only a small part of the emitted photons. Thus, the effect of details of Cherenkov emission distributions appears to be a second order effect.

\begin{figure}[tp]
\begin{center}
\includegraphics[scale=0.9]{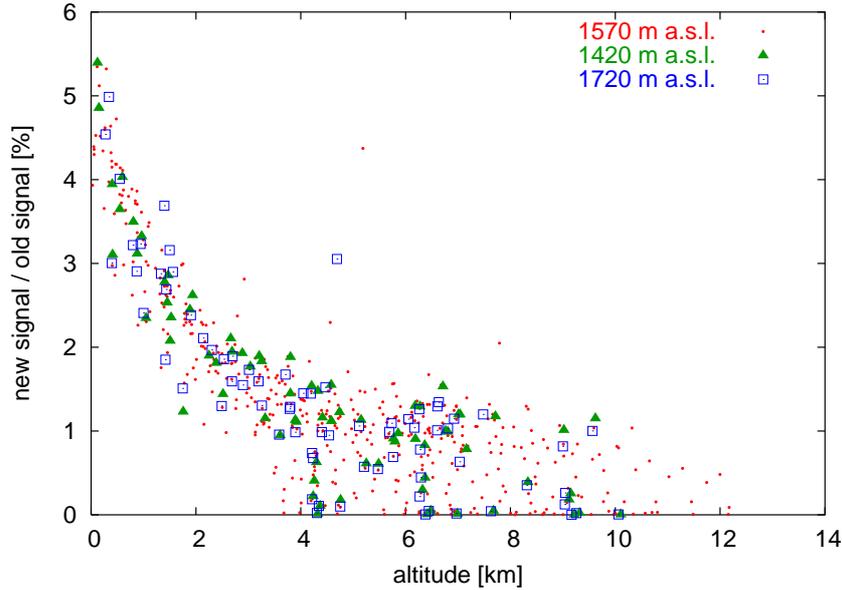}
\caption{\label {fig5bis} Comparison of contributions from scattered light, simulated for different locations of detectors above sea level.}
\end{center}
\end{figure}

All described above shower simulations were done for a detector placed at an altitude of 1570 meters above sea level, which is the default altitude in the \textit{Hybrid\_fadc} program and corresponds to altitude of HiRes detectors (1550 and 1593 m a.s.l.). It is also roughly the mean of the altitude range of fluorescence detector locations at the southern Pierre Auger Observatory (located between 1421 and 1719 m a.s.l.). It is necessary to check if the scattering contribution changes with altitude of the detector, since the air density, important for calculations of scattering, changes exponentially with altitude. It should be noted that Rayleigh scattering changes with air density, that is with altitude above sea level, while aerosol concentration, and consequently Mie scattering, changes independently according to local conditions and altitude above ground. Again, sets of simulations were made, with the detector placed 150 meters below and 150 meters above the default altitude, i.e. at 1420 and 1720 meters a.s.l. The results are shown in Fig. \ref{fig5bis}. Different simulation sets show no significant difference in the scattering contribution. It may not be surprising -- the difference of 300 meters in altitude translates to only 3\% change of air density (and in horizontal attenuation length for Rayleigh scattering), so no large changes in the scattering effect should be expected.

\begin{figure}[ht]
\begin{center}
\includegraphics[scale=0.5]{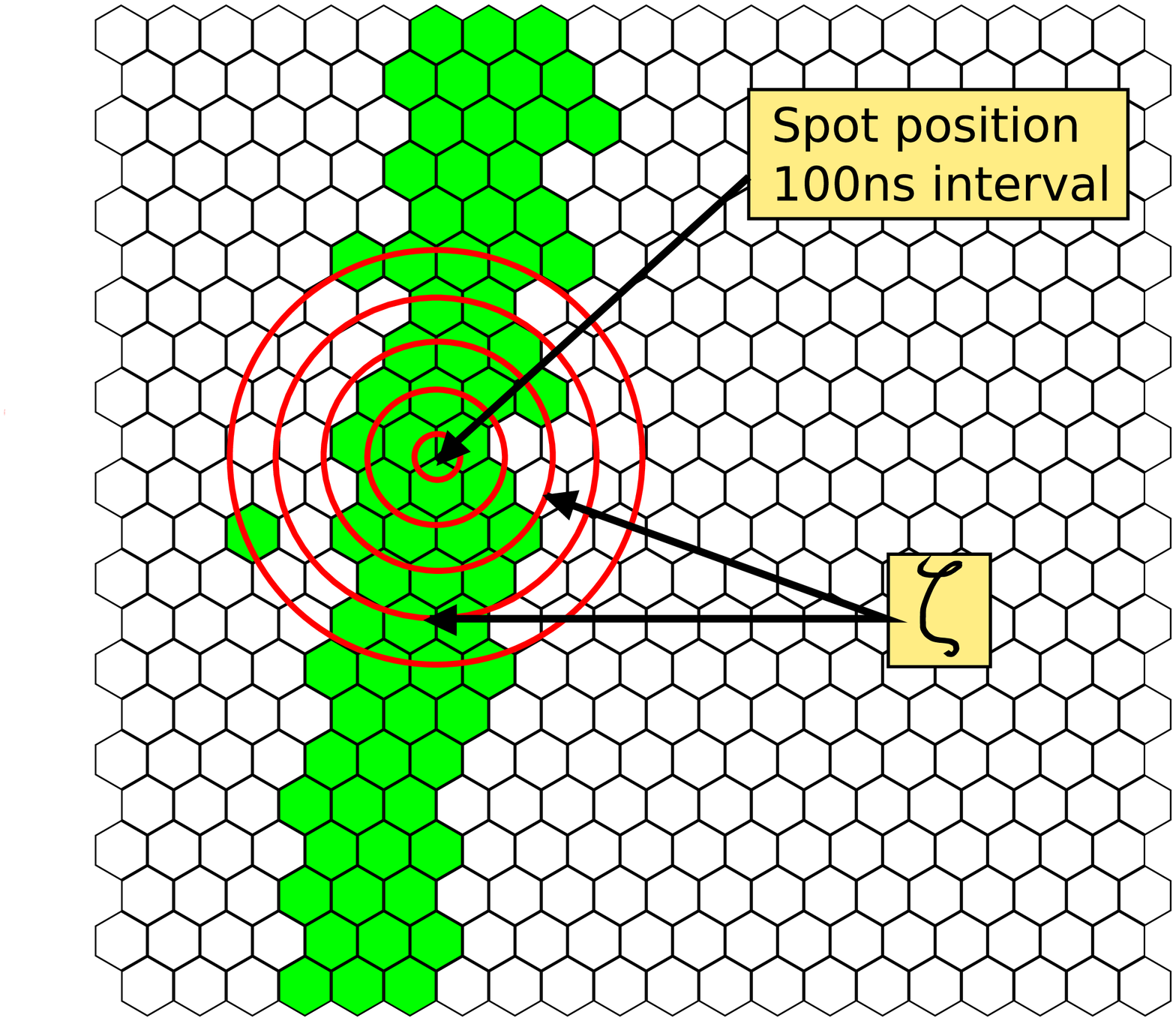}
\caption{\label {plamka} Schematic illustration of an air shower observation in a detector. The field of view of the detector is divided into hexagonal pixels, each observed by a photomultiplier. Light from air shower triggers pixels along the observed line of its propagation. The recorded signal is integrated within different angles $\zeta$ to find the best signal-to-noise ratio.}
\end{center}
\end{figure}

\begin{figure}[tp]
\begin{center}
\includegraphics[scale=0.9]{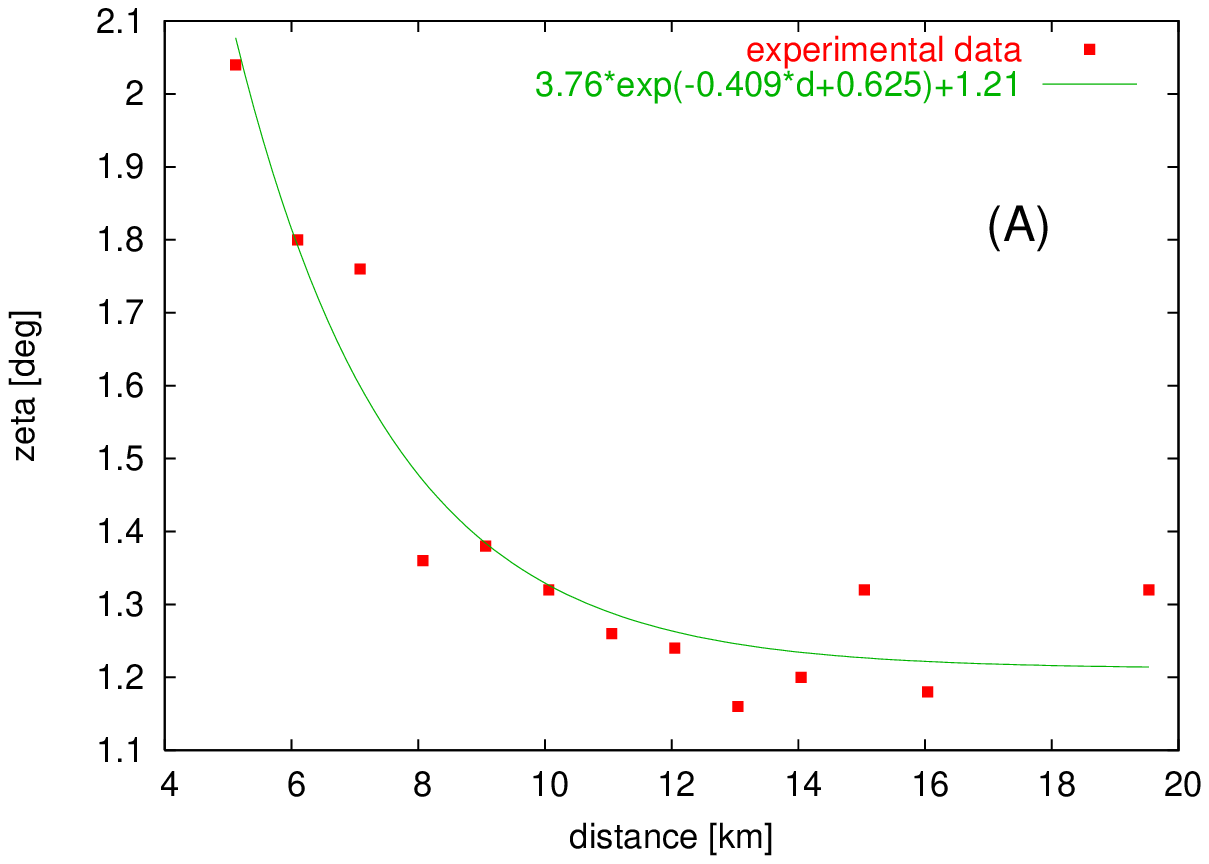}\\
\includegraphics[scale=0.9]{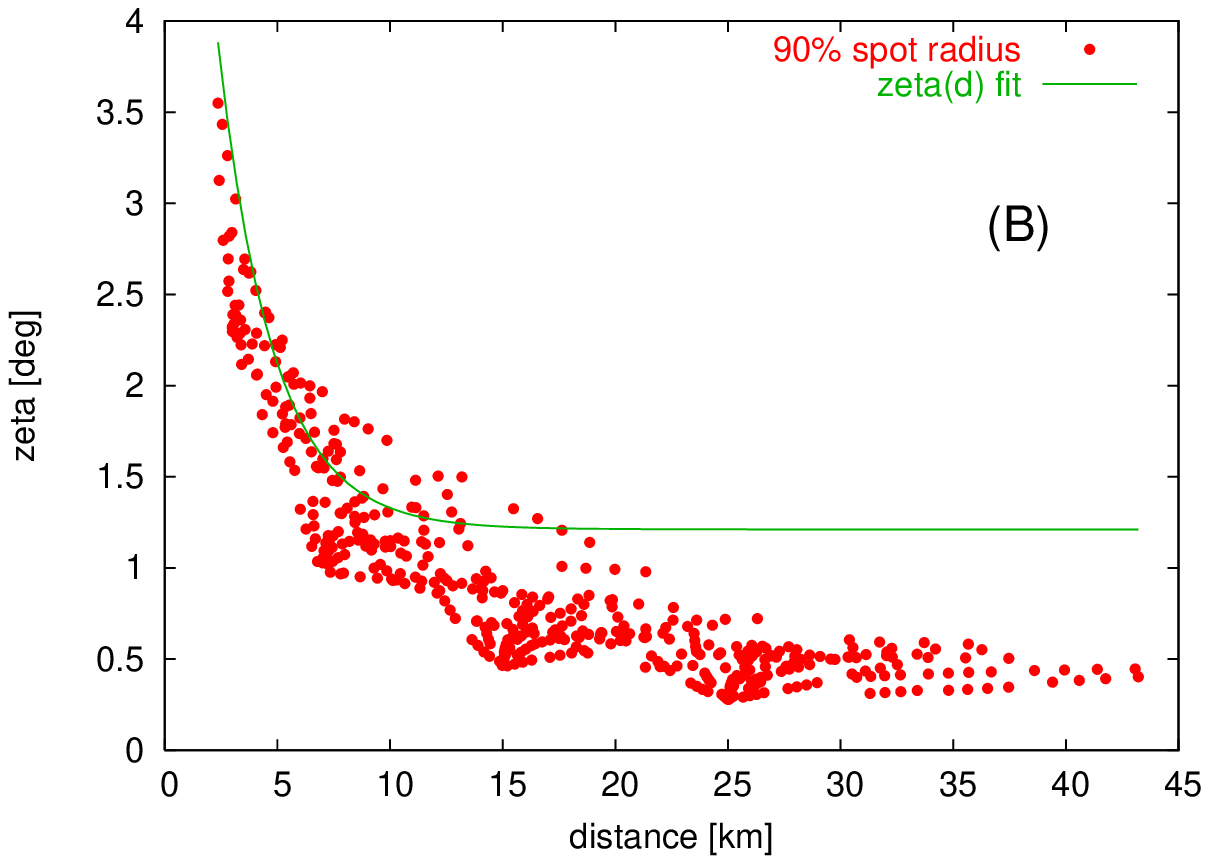}\\
\caption{\label {fig6} (A) Averaged values of $\zeta$ angle used in shower reconstruction (from Auger data \cite{desouza}) for different shower distances and an exponential fit to these points. (B) This $\zeta$ fit compared with the radius of image spot containing 90\% of shower light.}
\end{center}
\end{figure}

The results presented above demonstrate the contribution of the multiple scattering effect to the optical image of the shower defined as a circle containing 90 \% of the signal. In a real detector, the field of view is divided into pixels of a fixed size ($1.5^{\circ}$ in diameter in the Pierre Auger Observatory fluorescence detector \cite{abraham}, $1^{\circ}$ in the HiRes detector \cite{hires}). In the shower reconstruction procedure, the shower signal is obtained by summing signals from pixels located within some angle $\zeta$ from a current center of the shower image (Fig. \ref{plamka}).
The size of the $\zeta$ angle is chosen for each shower individually, to maximize the signal-to-noise ratio \cite{abraham}. On Fig. \ref{fig6}A shown is the dependence of the $\zeta$ angle on the distance from the shower to the detector, for showers recorded by the Auger Observatory \cite{desouza}. The exponential function fit to the data is compared to the size of the 90\% image spot in Fig. \ref{fig6}B. For the distant showers, the $\zeta$ angle is larger than the radius of the image spot. Detector properties, in particular pixellization of the field of view, are important here -- the $\zeta$ angle has to be always larger than the radius of a pixel. On the other hand, the spot size of the optical image of a shower depends on geometry and decreases with distance: being comparable to $\zeta$ angle for nearby showers, the image size becomes much smaller for distant showers.

As demonstrated in Fig. \ref{fig2}B, the angular distribution of scattered light is much wider than that of the direct light (i.e. wider than the ``shower'' curve). Therefore, the relative contribution of the scattered light to the recorded signal depends on the $\zeta$ angle. Since the goal of this work is to account for the scattering effect in analyses of data from real detectors, the dependence of the scattering contribution on $\zeta$ is essential.

\begin{figure}[t]
\begin{center}
\includegraphics[scale=0.9]{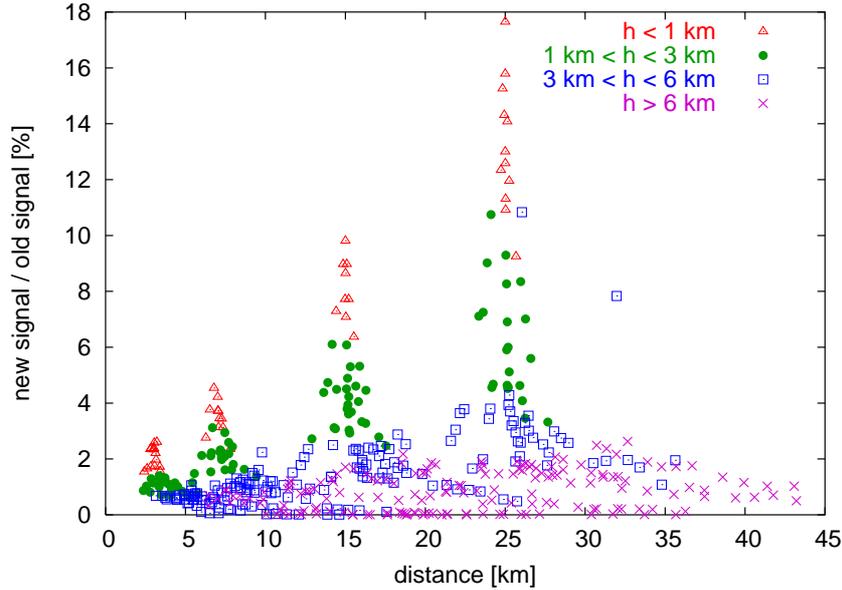}
\caption{\label {fig7} Contribution of new signal to shower signal versus shower-detector distance within light collecting angle $\zeta=1^{\circ}$.}
\end{center}
\end{figure}

\begin{figure}[tp]
\begin{center}
\includegraphics[scale=0.9]{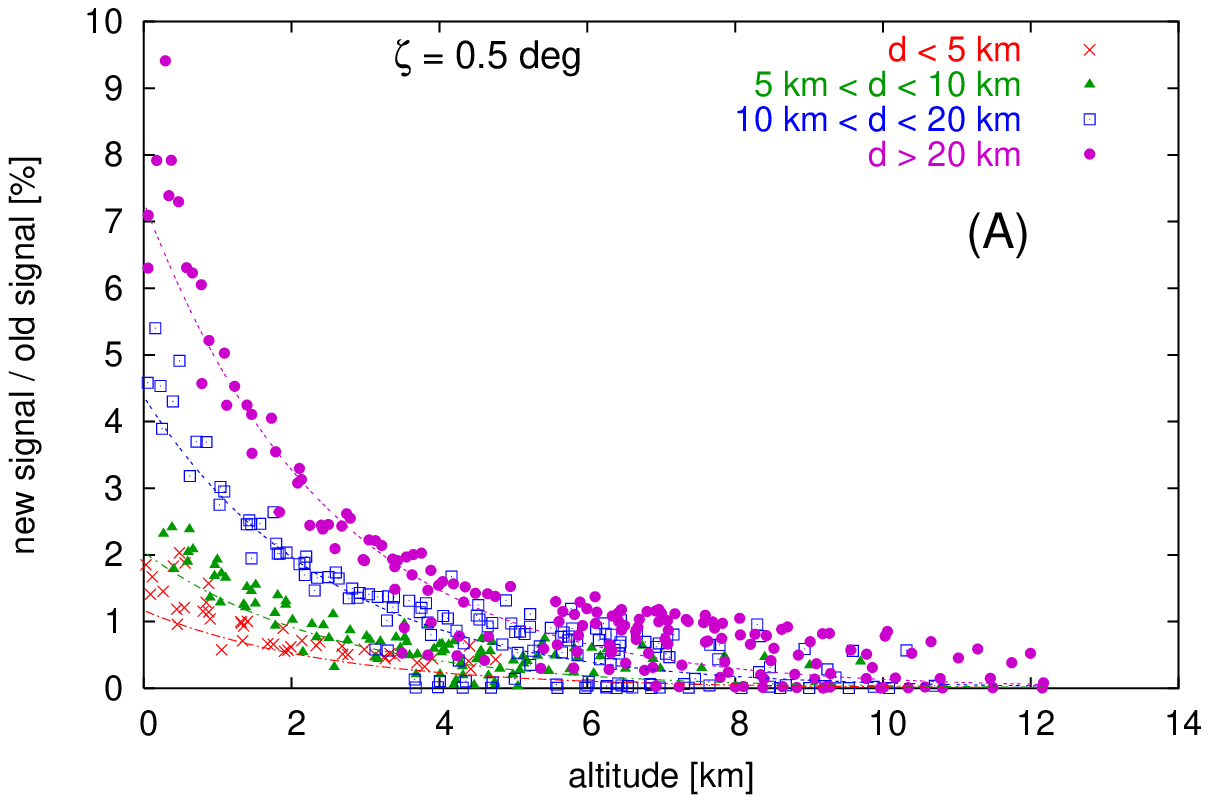}\\
\includegraphics[scale=0.9]{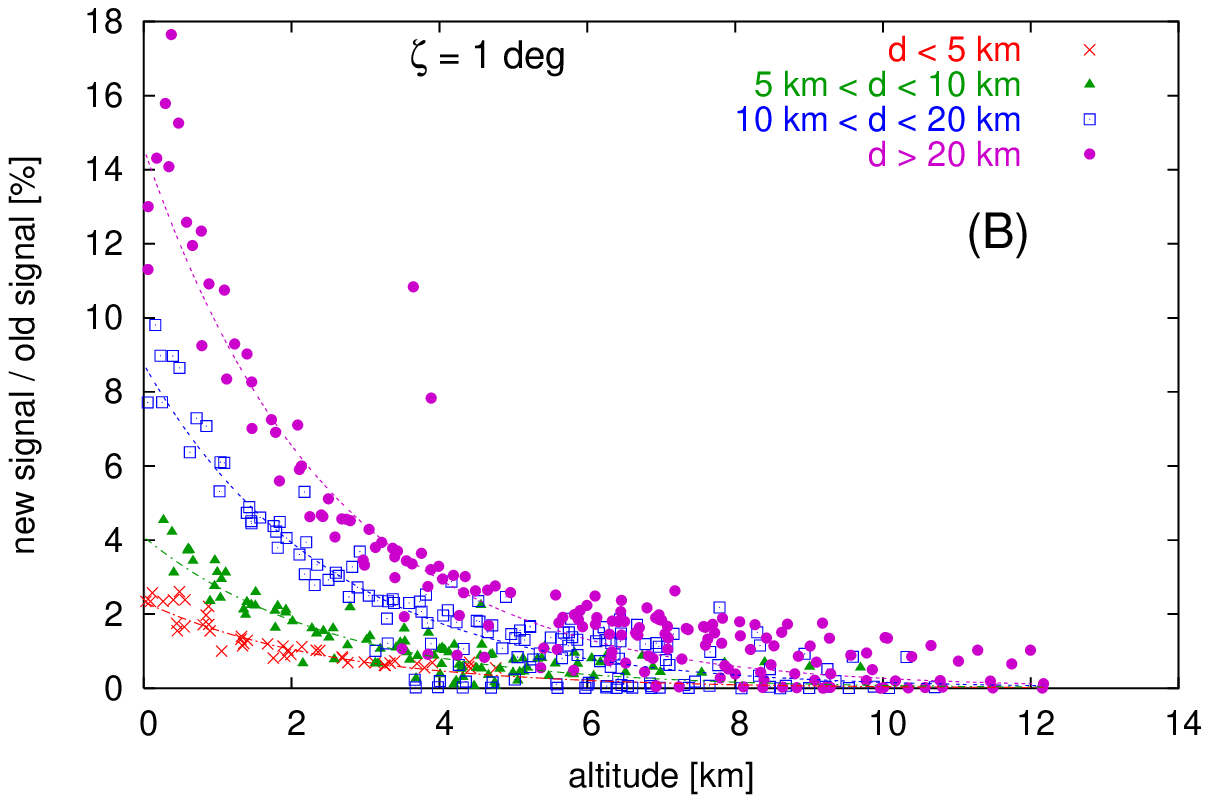}\\
\caption{\label {fig8} Contribution of the new signal to shower signal versus altitude above ground for different values of light collecting angle $\zeta$. The lines represent fits of Equation (\ref{fit}) for $d$ = 4, 7, 15 and 25 km.}
\end{center}
\end{figure}

For nearby showers, $\zeta$ is comparable to the 90\% spot radius. With $\zeta$ angle larger than 1$^{\circ}$, for distant showers, the light from solid angle much larger than shower image spot is recorded in the detector. This means that the contribution from multiple scattering to the recorded signal in the detector becomes larger for distant showers than for nearby ones. The relative contribution of the scattered signal calculated not for the 90\% spot radius (as in Fig. \ref{fig3}), but for various angles $\zeta$ is shown in Fig. \ref{fig7} and \ref{fig8}. The scattering contribution grows roughly linearly with increasing distance (this can be seen by comparing scales of vertical axes on Fig. \ref{fig8}). For typical $\zeta$ values of 1$^{\circ}$-1.5$^{\circ}$ used in shower reconstruction and for distant showers, the ``new signal'' from scattered fluorescence and multiply scattered Cherenkov photons exceeds 10\% of the direct shower signal at low altitudes. As shown on plots of Fig. \ref{fig8}, the character of the dependence on distance and altitude is similar for different $\zeta$ values, and the value of the scattering contribution scales approximately linearly with $\zeta$. On all plots of Fig. \ref{fig8} presented are simulated data points representing maximum of shower signal, and also at 250 g/cm$^{2}$ before and 250 g/cm$^{2}$ after the shower maximum. No significant separation of data points at different stages of shower development (i. e. different shower age) can be observed; all points at a given altitude and distance have comparable values of the scattering signal, so the scattering contribution appears to be independent of the shower age.

\begin{figure}[tp]
\begin{center}
\includegraphics[scale=0.9]{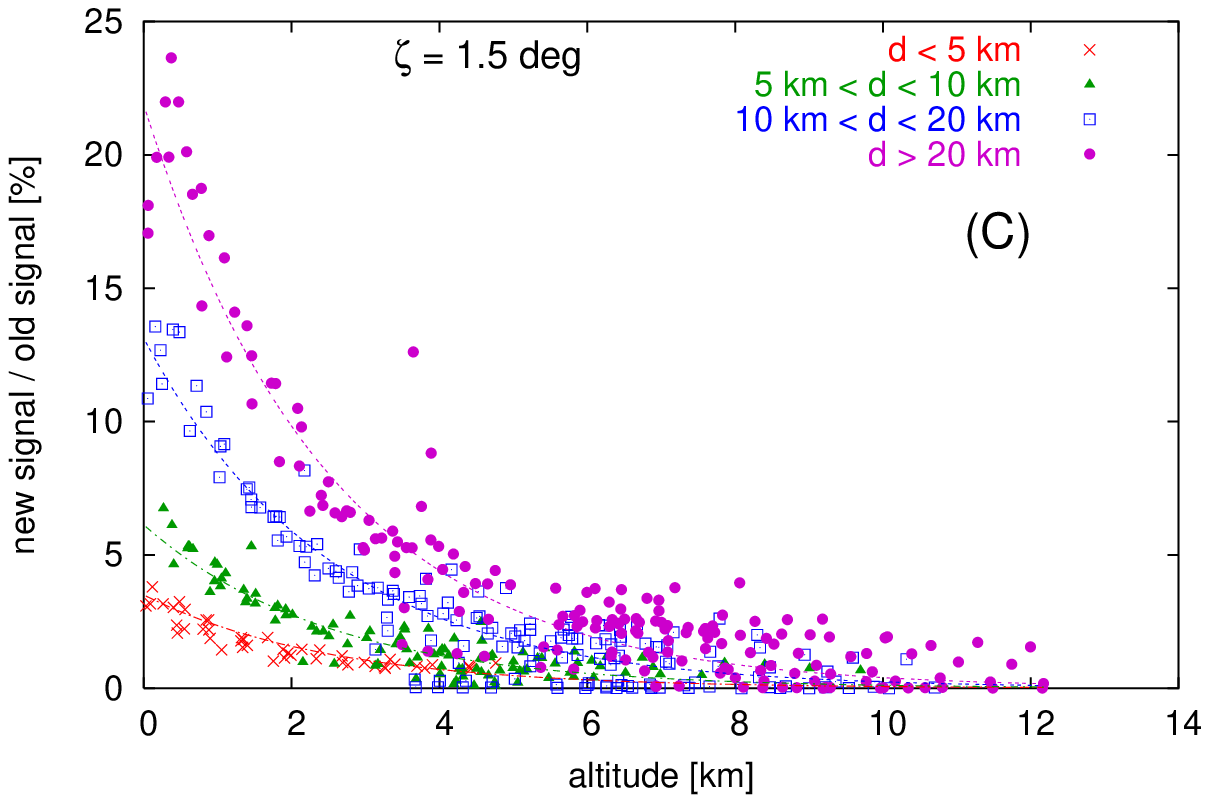}\\
\includegraphics[scale=0.9]{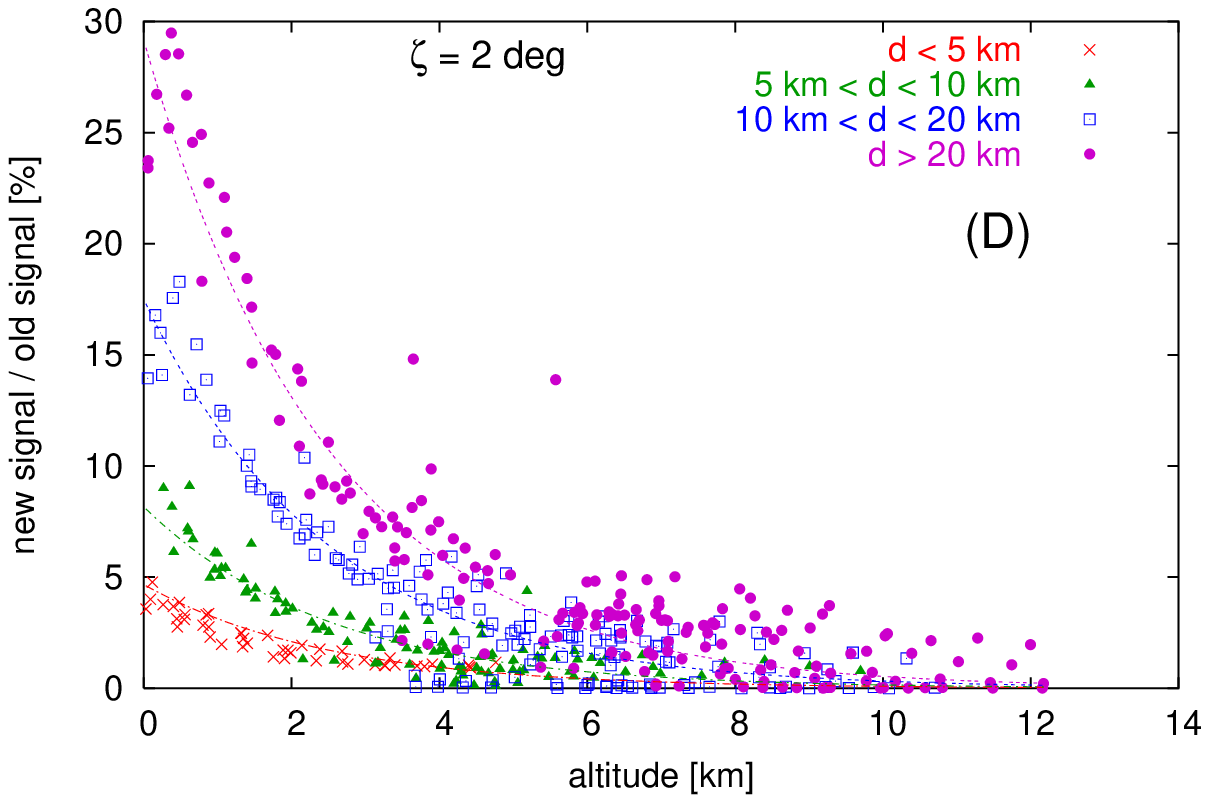}\\
Fig. \ref {fig8} continued.
\end{center}
\end{figure}

Another variable that must be taken into account is the amount of aerosols in the air. All previous calculations were done with one value of aerosol concentration, equivalent to horizontal attenuation length (at ground level) for Mie scattering of $\Lambda_{M}$=9.59 km, which is the default value of the ``Hybrid\_fadc'' program. For the Rayleigh scattering at the default detector altitude (1570 m a.s.l.) the horizontal attenuation length for the light wavelength of 361 nm is $\Lambda_{R}$=18.77 km. Therefore, the total horizontal attenuation length ($1/\Lambda_{T} = 1/\Lambda_{R} + 1/\Lambda_{M}$) is $\Lambda_{T}$= 6.347 km. $\Lambda_{R}$ and $\Lambda_{M}$ cannot be measured separately. $\Lambda_{R}$ depends on the air density only, so it is well known. $\Lambda_{M}$ can be obtained only by measuring $\Lambda_{T}$, so the total attenuation length $\Lambda_{T}$ can be used as a parameter in study of aerosol concentration variations. In order to check the influence of amount of aerosols on the contribution of multiple scattering, a set of simulations was done:

\begin{itemize}
\item total horizontal attenuation length $\Lambda_{T}$: 9.487, 11.827, 15.699 km (corresponding to 50, 30 and 10 percent of ``Hybrid\_fadc'' default aerosol concentration);
\item shower energy: $10^{19}$eV;
\item shower core distance: 3, 7, 15, 25 km;
\item ${\psi}$ angle: 30, 50, 70, 90, 110, 130, 150 degrees for vertical SDP;
\item SDP inclination: 30, 45, 60, 70 degrees with ${\psi}=90^{\circ}$.
\end{itemize}

\begin{figure}[tp]
\begin{center}
\includegraphics[scale=0.9]{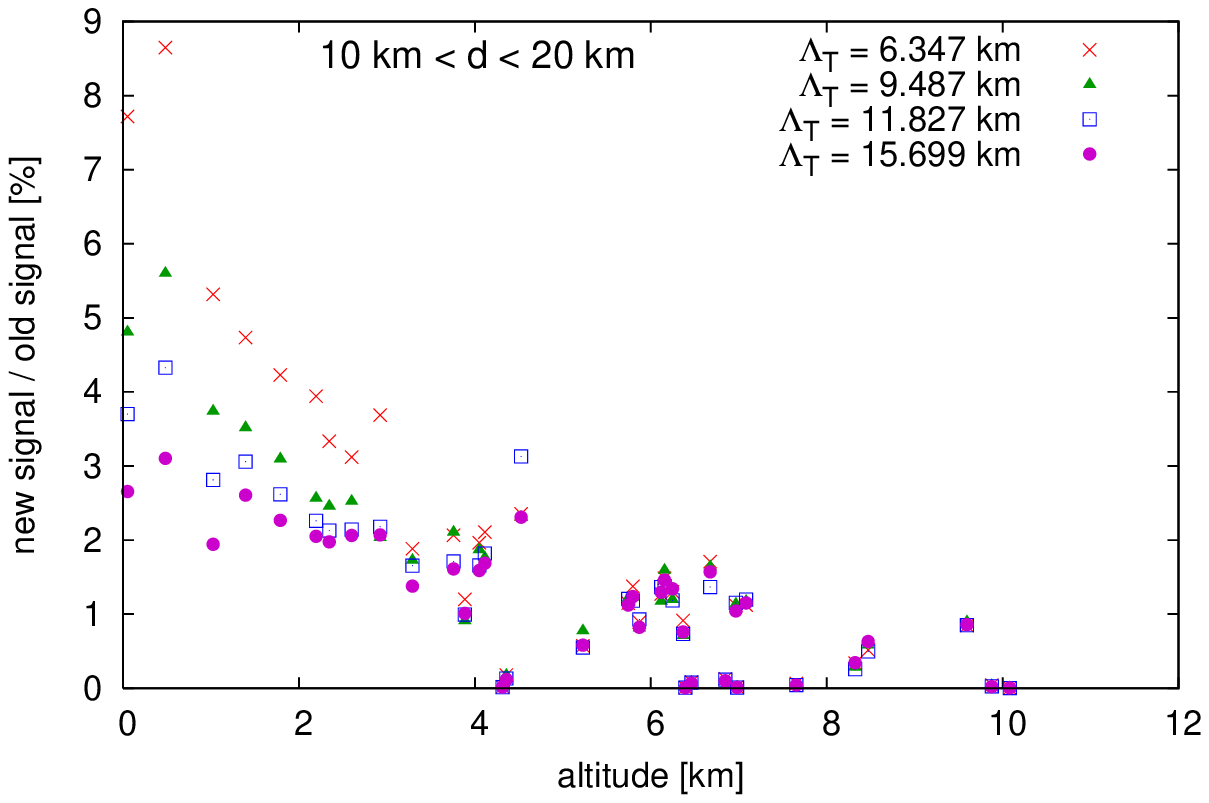}\\
\includegraphics[scale=0.9]{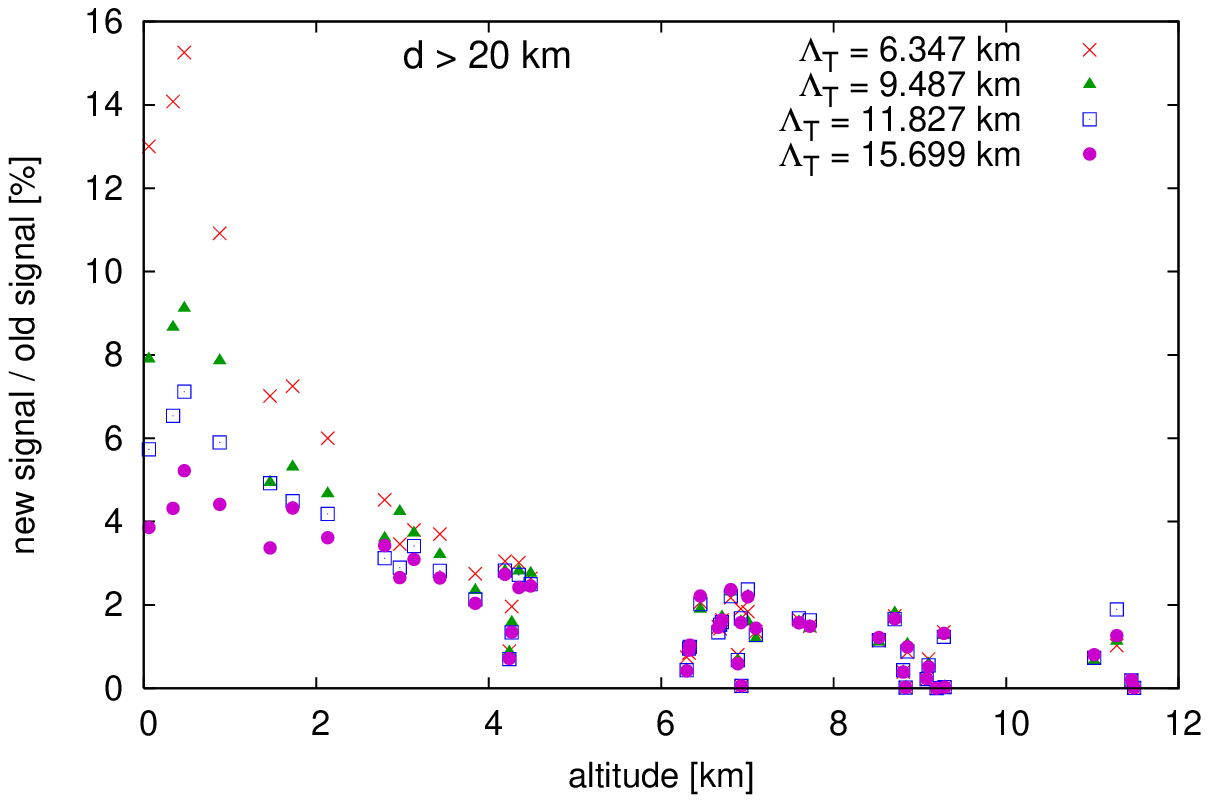}\\
\caption{\label {od1} Contribution of the new signal to shower signal versus altitude above ground for different aerosol concentrations characterized by different values of total horizontal attenuation length $\Lambda_{T}$. The contributions are integrated within the angle $\zeta=1^{\circ}$. Shown are points at selected distances from the detector.}
\end{center}
\end{figure}

Examples of results from these simulations are shown on Fig. \ref{od1}. It can be observed that the contribution of multiple scattering varies strongly with aerosol concentration. Especially for low altitudes above the ground, a dependence is seen: for higher aerosol concentration (smaller $\Lambda_{T}$) the scattering contribution is higher.

\begin{figure}[tp]
\begin{center}
\includegraphics[scale=0.9]{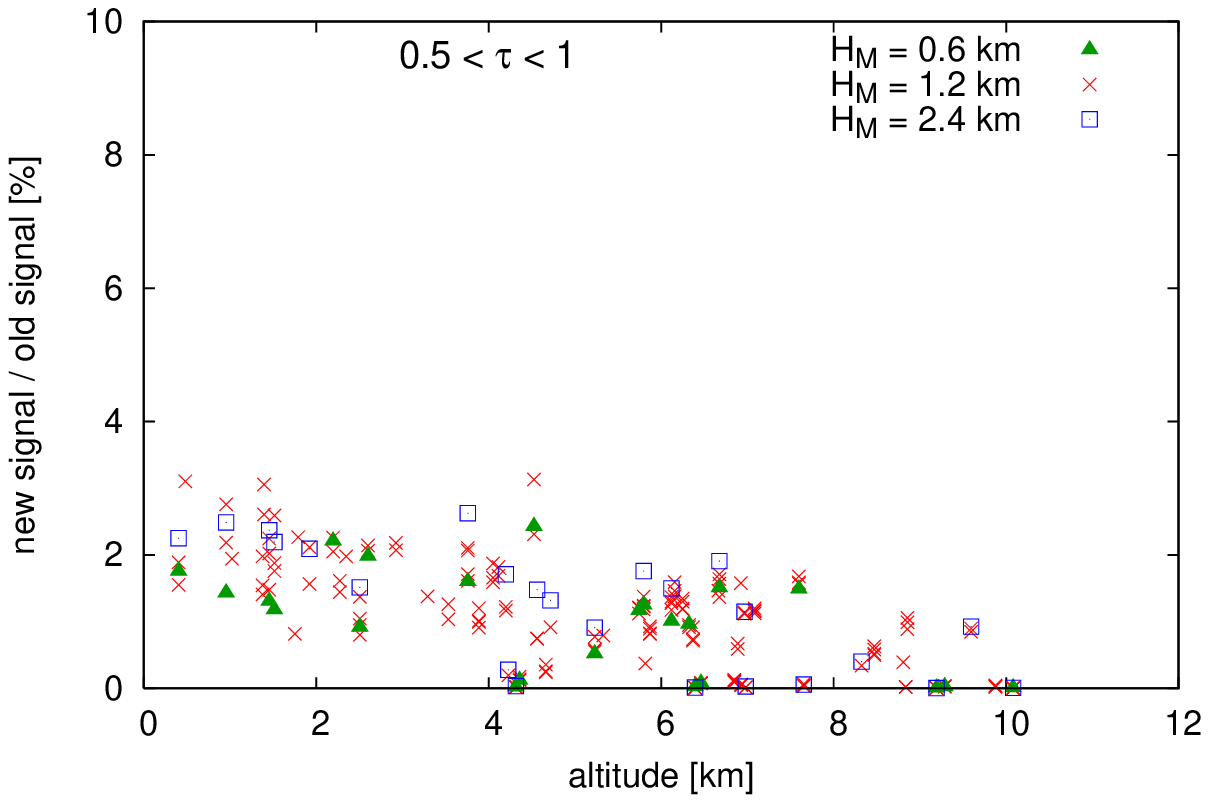}\\
\includegraphics[scale=0.9]{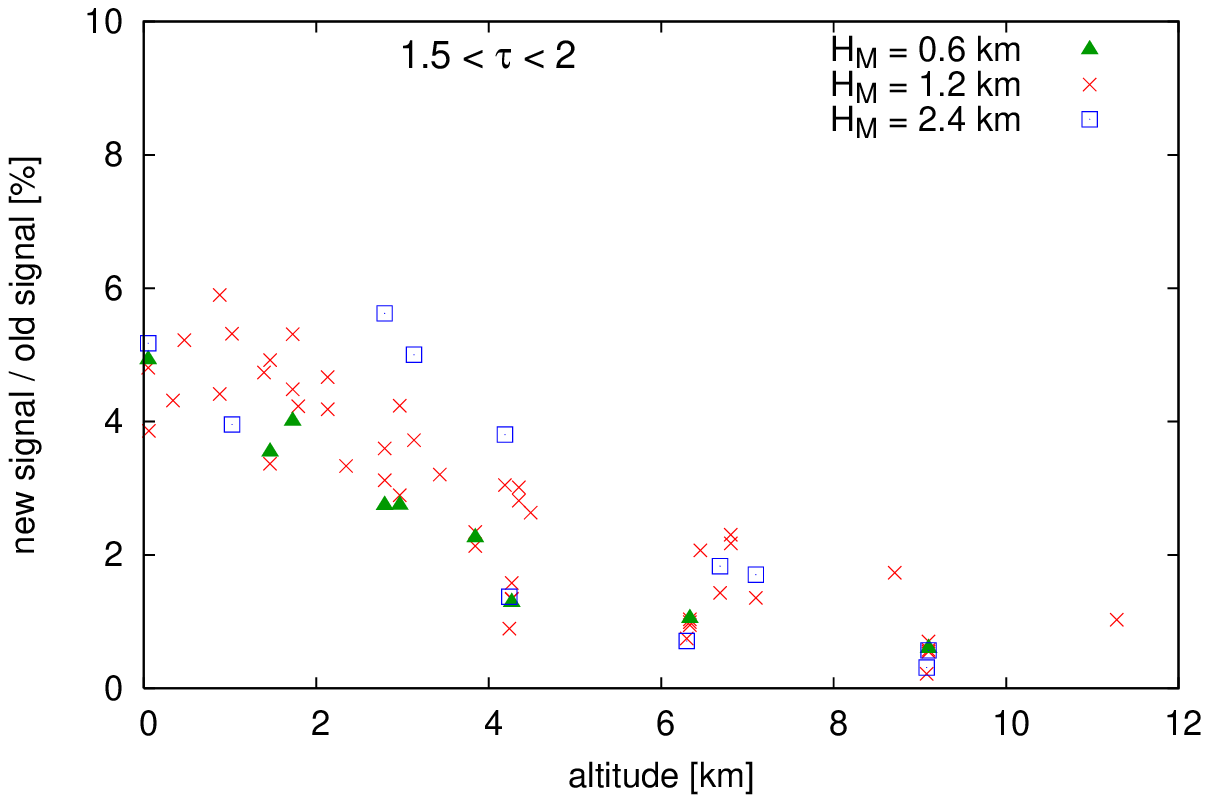}\\
\caption{\label {od2} Comparison of the new signal for different values of aerosol scale height H$_{\mathrm{M}}$. On the plots are shown points with similar values of optical depth. The contributions are integrated within the angle $\zeta=1^{\circ}$.}
\end{center}
\end{figure}

To describe fully the concentration of aerosols in the whole volume of air, not only $\Lambda_{M}$ must be known, but also the vertical distribution of aerosol particles. Generally it is assumed that the concentration of aerosols falls down exponentially with altitude, with a scale height of 1.2 km. This distribution was used in all multiple scattering simulations discussed above. In order to see if a change of aerosol scale height influences the final results, a set of simulations was made:

\begin{itemize}
\item aerosol scale height H$_{M}$: 0.6, 2.4 km:
\item $\Lambda_{T}$ = 9.487 km;
\item energy $10^{19}$eV;
\item core distance: 3, 7, 15, 25 km;
\item ${\psi}$ angle: 30, 50, 70, 90, 110, 130, 150 degrees for vertical SDP.
\end{itemize}

Results from these simulations are shown on Fig. \ref{od2}. The data points were selected so that they all have comparable values of optical depth $\tau$ for the line of sight from the emission point at the shower to the detector. The optical depth was selected because it describes both geometrical distance and scattering properties of the path from the air shower to the detector. On the plots compared are the multiple scattering contributions for different aerosol scale heights. It can be observed that even large differences of aerosol scale height (within a factor of 4), don't make a significant change -- no regular dependence on the scale height is seen. The contributions of multiple scattering for the same optical depth and altitude show some variation (1-2\%). With the aerosol scale height changed, the scattering contributions are still within the range of this variation. Therefore a conclusion can be made that variation of scale height of aerosols doesn't need to be separately accounted for.

\section{Parameterization of scattering contribution}

All observations made during the review of simulation results allow one to make a parameterization of the multiple scattering contribution. This contribution is calculated as: $$M = \dfrac{\mathrm{new \; signal}}{\mathrm{old \; signal}} [\%].$$ The old signal is the sum of direct fluorescence, direct Cherenkov and singly scattered Cherenkov light. The new signal is the sum of scattered (singly and multiply) fluorescence light and multiply scattered Cherenkov light. For one value of horizontal attenuation length $\Lambda_{T}$, $M$ may be parameterized as a linear function of the $\zeta$ angle (in degrees) and of the shower-detector distance $d$ and an exponential function of altitude above ground $h$ (both in kilometers)\footnote{We note that two geometrical variables are needed to make the parameterization: we chose $d$ and $h$}:
\begin{equation} \label{fit}
M=A \zeta d \exp(-\frac{h}{B})
\end{equation}
A fit of this function with three independent variables ($\zeta$,$d$,$h$) and two parameters ($A,B$) was made. Data from simulations for $\Lambda_{T}$=6.347 km (at 361 nm) were used, representing shower image at their maxima and points 250 g/cm$^{2}$ higher and lower (if above ground), in the range of the $\zeta$ angle between 0.1$^{\circ}$ and 5$^{\circ}$ with a step of $0.1^{\circ}$. The resulting parameters of the fit are: $A=0.5830\pm0.0011 \%/deg, \hspace{5mm} B=2.4986\pm0.0062 \mathrm{km},$ when $\zeta$ is in degrees, $d$ and $h$ in kilometers.

In order to confirm that the dependence on $\zeta$ and \textit{d} is linear, an additional fit to a following function was made:
$$M=A \zeta^{C} d^{D} \exp(-\frac{h}{B}).$$
The resulting values of the exponents: $C=0.986\pm0.007, \hspace{5mm} D=0.993\pm0.006$ show that treating the scattering contribution as a linear function of $\zeta$ and \textit{d} is an acceptable simplification.

\begin{figure}[t]
\begin{center}
\includegraphics[scale=0.9]{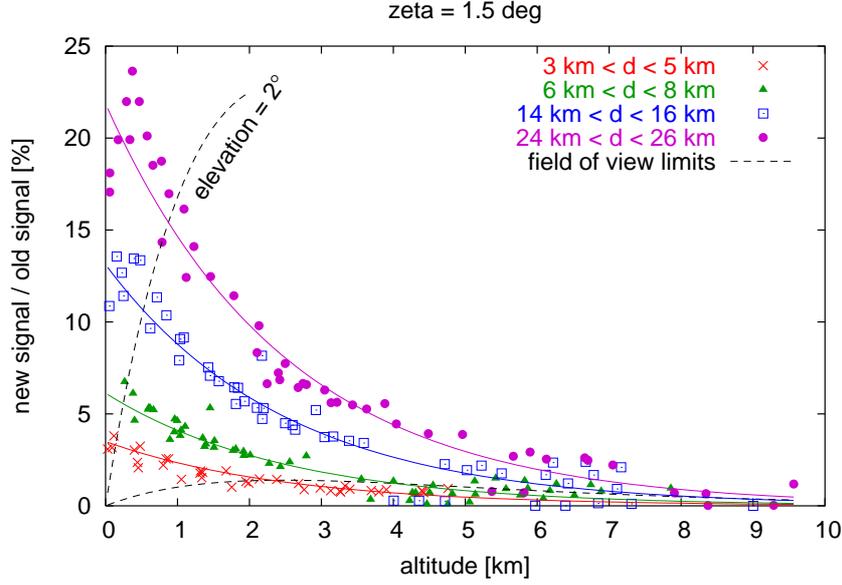}\\
\caption{\label {fig9} Contribution of the new signal to shower signal versus altitude above ground for selected shower-detector distances for $\zeta=1.5^{\circ}$. Solid lines represent fits to function (\ref{fit}) to all simulation points for $d$ = 4, 7, 15 and 25 km. Dashed lines show field of view limits of Auger fluorescence telescopes (2 and 30 degrees above the horizon).}
\end{center}
\end{figure}

Fits of Eq. \ref{fit} to the simulation results are shown in Figure \ref{fig8}. To better show the agreement with the data points, a subset of the data in smaller intervals of distance $d$ is shown in Figure \ref{fig9}. The fits are shown for $d$ = 4, 7, 15 and 25 km, with data points corresponding to distances within $\pm1$ km from these values. Equation \ref{fit} fairly well represents the contribution of the new signal to shower signal in a fluorescence detector.

The scattering contribution strongly depends on altitude of the shower front above the ground, and consequently -- on the elevation angle of the line of sight. If the field of view of a detector is limited, it may limit the range of the scattering contribution values that are really observed. As an example, the approximate limits of the field of view of the Pierre Auger Observatory fluorescence detectors (between $2^{\circ}$ and $30^{\circ}$ above the horizon) are marked in Fig. \ref{fig9} by the dashed lines. This means that only points located to the right of the ``elevation=2$^{\circ}$'' line represent contributions which are relevant for the Auger detectors. Nevertheless, the multiple scattering contribution to the recorded signal may exceed 10\%.

\begin{figure}[tp]
\begin{center}
\includegraphics[scale=0.9]{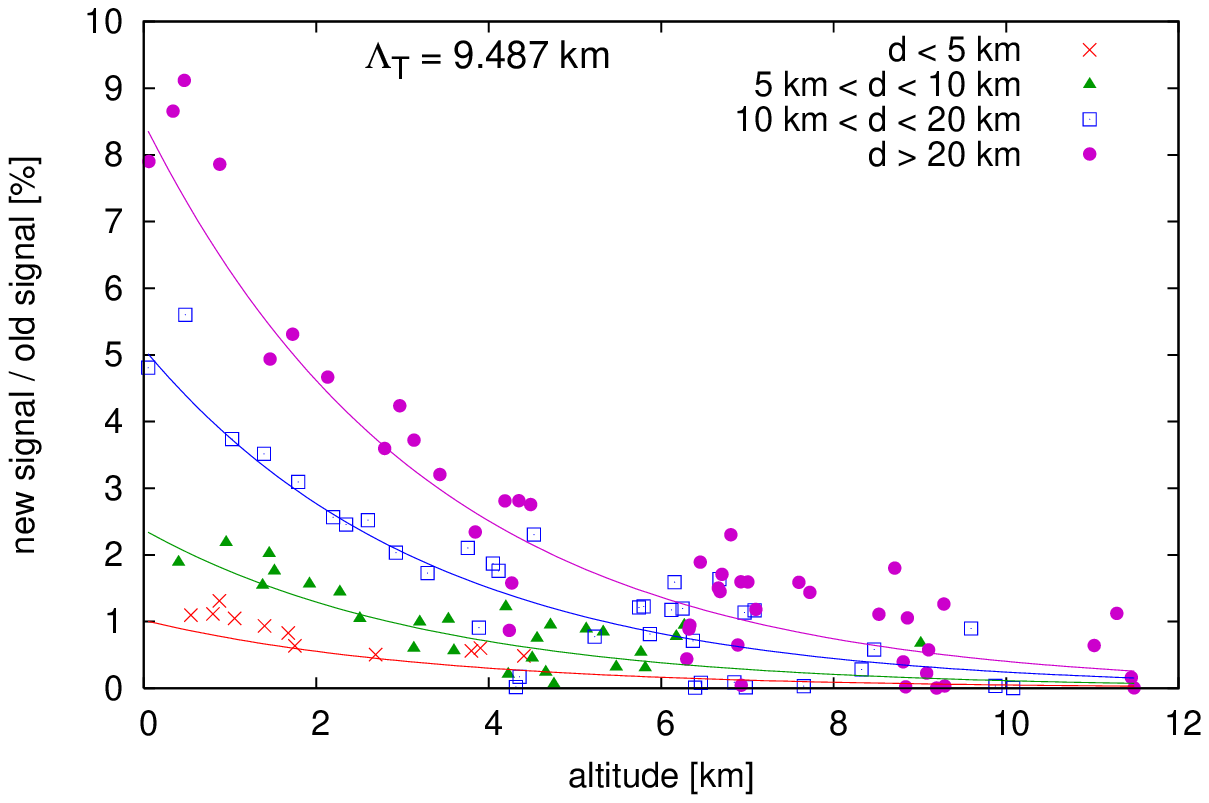}\\
\includegraphics[scale=0.9]{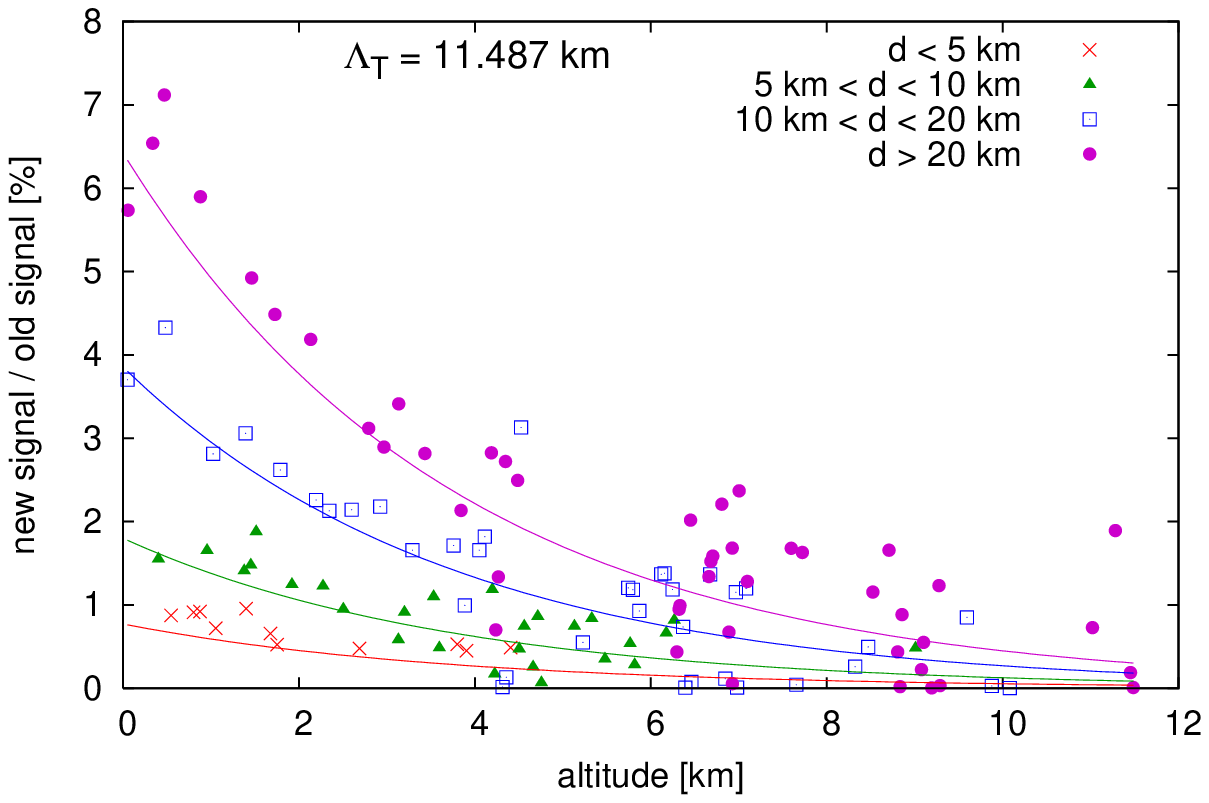}\\
\caption{\label {od3} Contribution of the new signal for different values of the total horizontal attenuation length. The contributions are integrated within the angle $\zeta=1^{\circ}$. The lines represent fits of Eq. \ref{fit}.}
\end{center}
\end{figure}

For other values of the horizontal attenuation length, similar fits can be performed. The results for two of the other $\Lambda_{T}$ values are shown on Fig. \ref{od3}. As it can be observed, Equation \ref{fit} describes fairly well the scattering contribution for different aerosol concentration, especially for relatively large values of the contribution.
\begin{figure}[t]
\begin{center}
\includegraphics[scale=0.9]{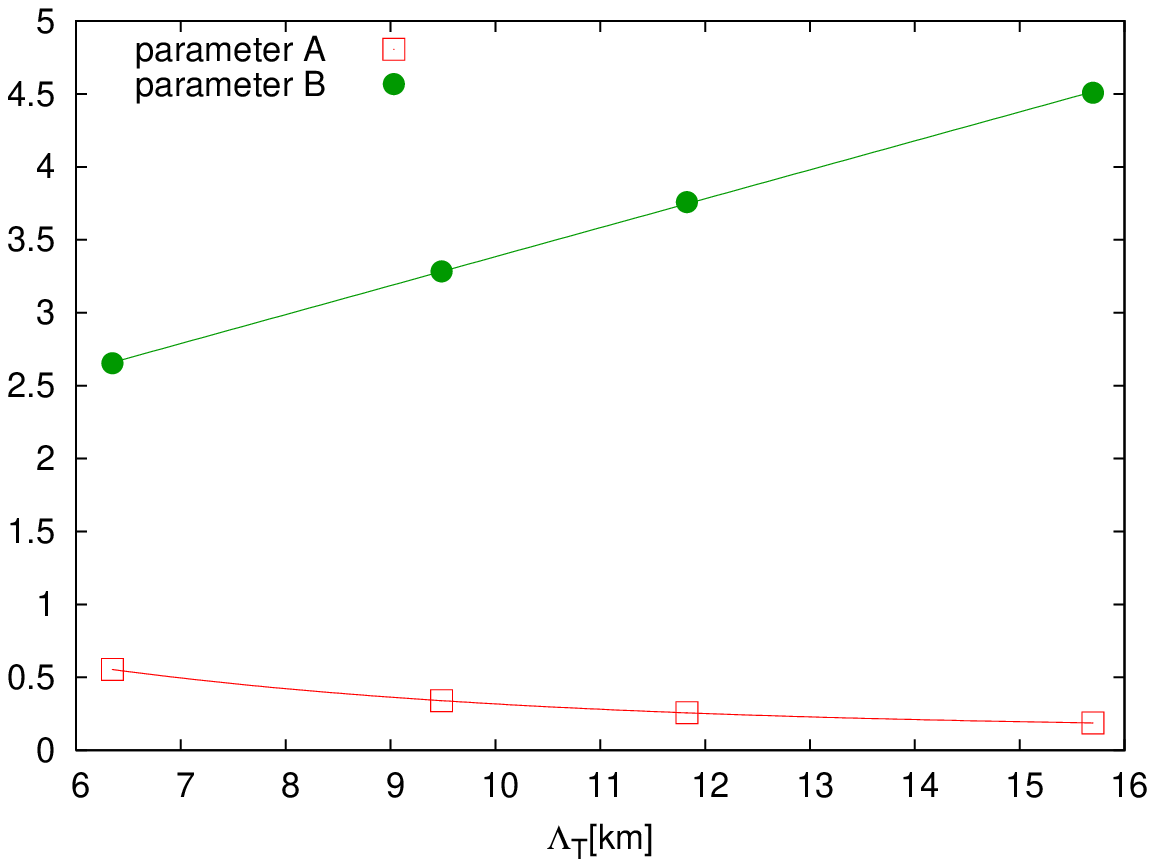}\\
\caption{\label {od4} Dependence of the $A$ and $B$ parameters (Eq. \ref{fit}) on the total horizontal attenuation length.}
\end{center}
\end{figure}
However, the A and B parameters of Equation \ref{fit} depend on the horizontal attenuation length (see fig. \ref{od4}). Therefore, in order to use this function as parameterization of multiple scattering contribution, the A and B parameters must be described as functions of $\Lambda_{T}$. And so, as it is shown in Fig. \ref{od4}, parameter A can be approximated by an exponential function, and B -- by a linear function. When presenting the Equation \ref{fit} in this form, the following values are obtained from the fit:

\begin{equation} \label{param1}
\begin{split}
A =a_{1}& \exp(-\Lambda_{T}/a_{2}) + a_{3},\\
\mbox{where } a_{1}& =1.774\pm0.033 \%,\\
\hspace{11mm} a_{2}& =4.365\pm0.064 \mathrm{km},\\
\hspace{11mm} a_{3}& =0.1387\pm0.0025 \%;\\
B =b_{1}& \Lambda_{T} + b_{2},\\
\mbox{where } b_{1}& =0.1976\pm0.0035,\\
\hspace{11mm} b_{2}& =1.402\pm0.029 \mathrm{km}.
\end{split}
\end{equation}

With these expressions for the $A$ and $B$ parameters, the parameterization of the multiple scattering contribution by Eq. \ref{fit} and Eq. \ref{param1} is complete. This contribution is given as a function of the signal integration angle $\zeta$, the shower-detector distance $d$, the altitude of the current shower position $h$ and the total horizontal attenuation length $\Lambda_{T}$.

\begin{figure}[t]
\begin{center}
\includegraphics[scale=0.9]{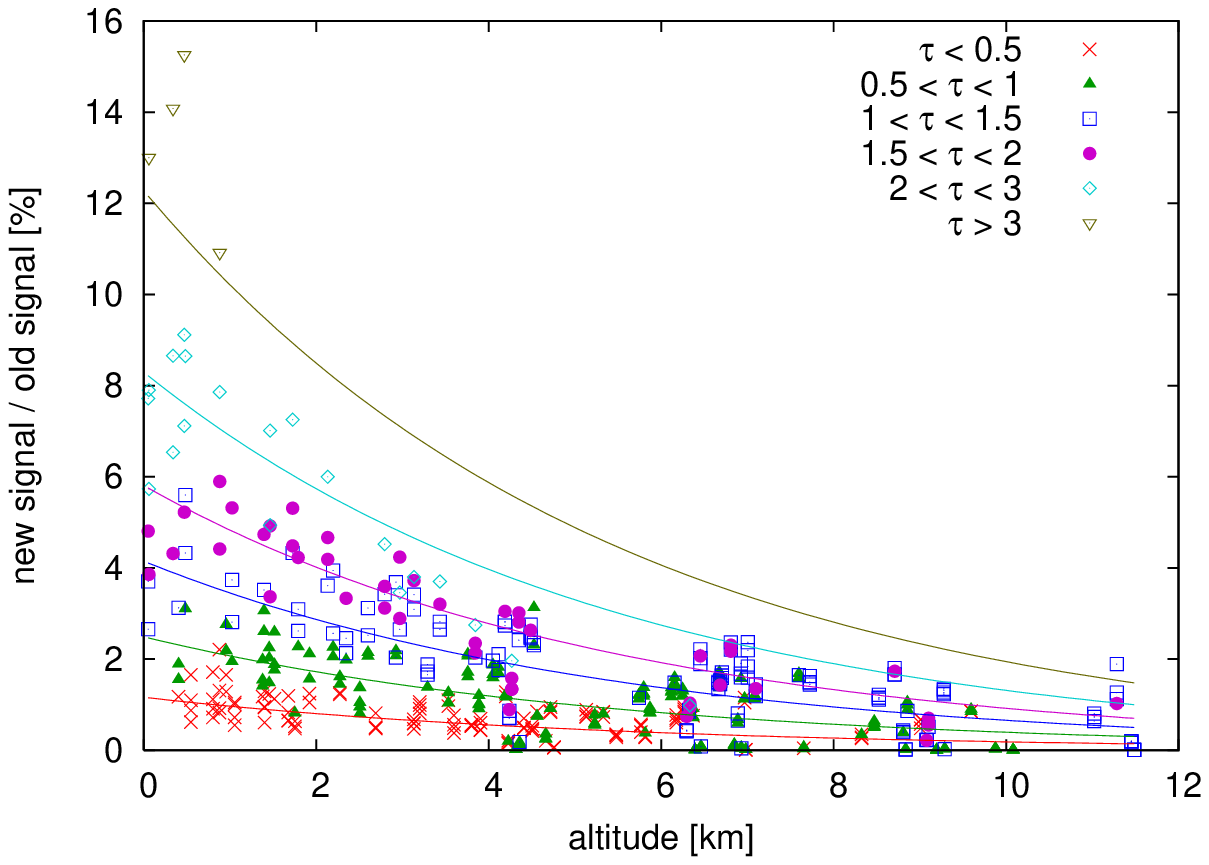}\\
\caption{\label {od5} Contribution of the new signal for all values of horizontal attenuation length. Points are grouped by their value of optical depth for the shower-detector distance. The lines are fits of Eq. \ref{fit2}, drawn for the mean value of respective $\tau$ ranges.}
\end{center}
\end{figure}

For different aerosol concentrations in the air, the scattering contributions are different, even for the same geometrical distance and altitude of the shower front. Therefore it can be concluded that a parameter independent of individual geometry could be more convenient. The optical depth $\tau$ is a parameter that characterizes the distance between two points, with respect to the atmospheric scattering on the way between these points. Indeed, an observation can be made: if the scattering contributions for different $\Lambda_{T}$ are grouped not by their geometrical distance $d$ but by optical depth $\tau$, the scattering contribution values are similar for the same altitudes. Therefore, the function to which the data is fitted, may be the following:
\begin{equation} \label{fit2}
M= F \zeta \tau \exp(-\frac{h}{G})
\end{equation}
$\tau$ is calculated for 361 nm wavelength.
A fit to the function \ref{fit2} gives the following values of the parameters:
\begin{equation} \label{param2}
F = 3.3179 \pm 0.0078 \%/deg, \hspace{5mm} G = 5.426 \pm 0.027 \mathrm{km}.
\end{equation}
As it is shown in Fig. \ref{od5}, a function of this form also gives a reasonably good fit to the results from simulations. The fitted function describes the results from simulations with accuracy of about 1-2 \%, which is sufficient for applying this correction. Statistical fluctuations of results from individual simulations don't allow a more accurate parameterization. In this form the parameterization of multiple scattering contribution is a function of the $\zeta$ angle, optical depth $\tau$ along the shower-detector line and the geometrical altitude of shower $h$.

\begin{figure}[tp]
\begin{center}
\includegraphics[scale=0.95]{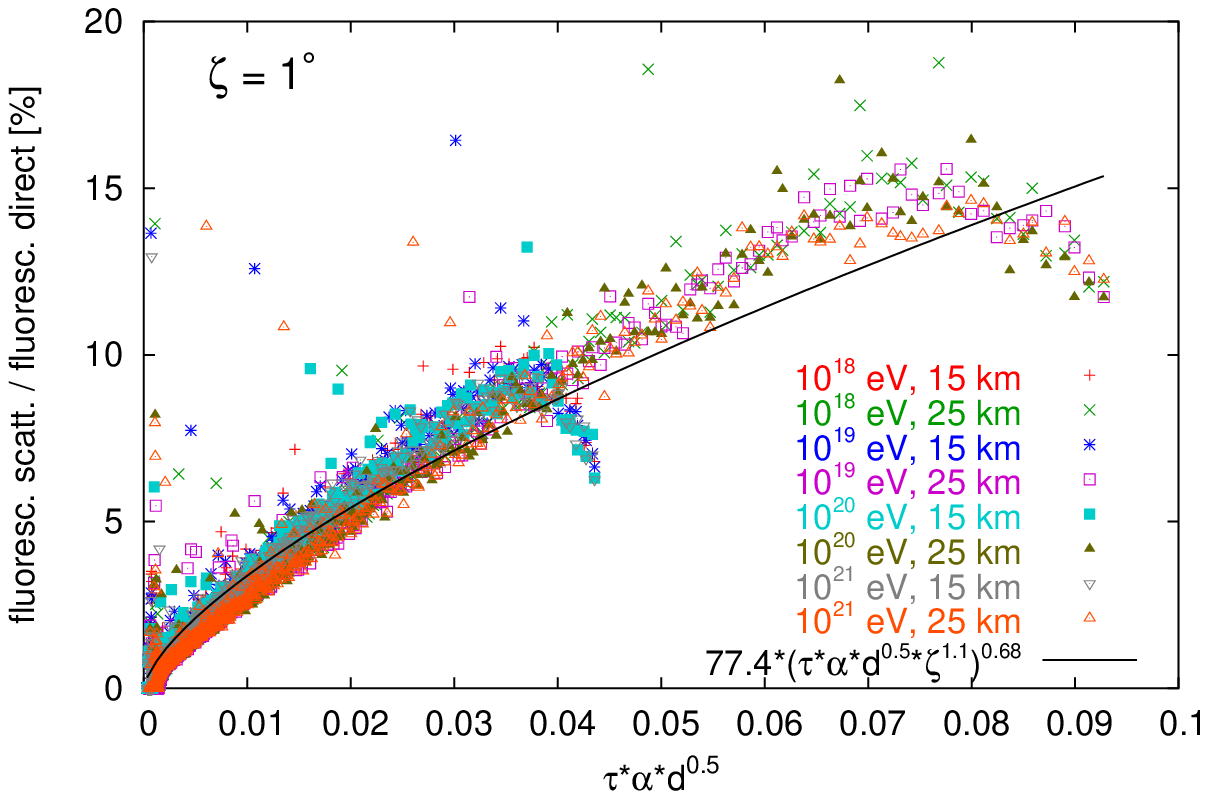}\\
\includegraphics[scale=0.95]{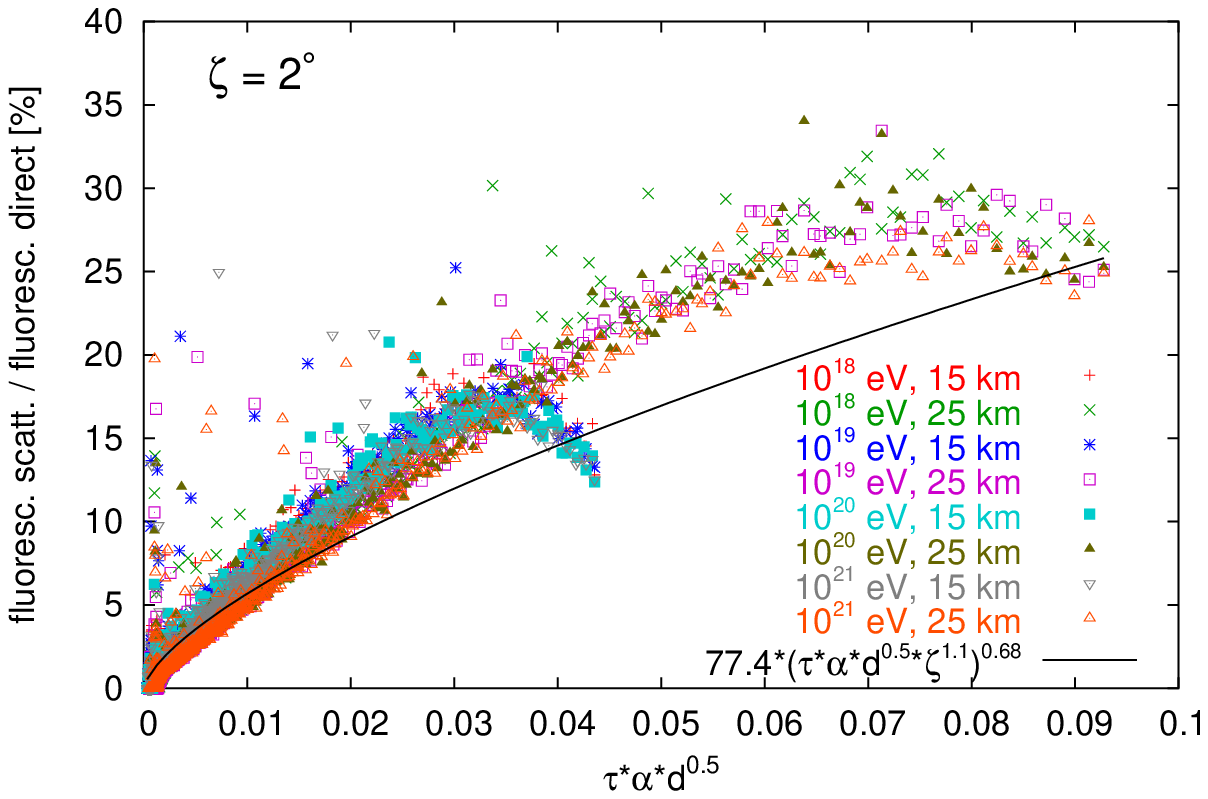}\\
\caption{\label {fig11} Comparison of fluorescence scattering from our set of simulations (data points) with results of \cite{roberts} for $\zeta=1^{\circ}$ and $\zeta=2^{\circ}$. The lines represent the parameterization given in Ref. \cite{roberts}.}
\end{center}
\end{figure}

The results of our simulations are compared to results of other studies \cite{roberts,giller}, wherever possible. In Ref. \cite{roberts} the multiple scattering of fluorescence light for vertical showers was studied, assuming a uniform fluorescence light intensity along the shower path. To make the comparison, scattered fluorescence light only from the vertical showers from our simulations is plotted in Fig. \ref{fig11} as a function of quantity used in \cite{roberts}: $\tau*\alpha*\sqrt{d}$, where $\tau$ is the optical depth between the shower and the detector, $\alpha$ is the total scattering coefficient at the point of emission [$m^{-1}$], $d$ is the shower-detector distance [m]. Results for $\zeta=1^{\circ}$ and $\zeta=2^{\circ}$ are compared. Although the methods of simulation in both studies are different, the results appear to be in a reasonable agreement, especially for small $\zeta$. One can observe on Fig. \ref{fig11} that the contribution from scattering grows as a function of $\tau*\alpha*\sqrt{R}$, except for the final phase of shower development. This can be explained by proximity to the ground level -- larger part of light produced by shower particles is absorbed at the ground before scattering in air can occur, thus decreasing the scattering contribution.

\begin{figure}[tp]
\begin{center}
\includegraphics[scale=0.95]{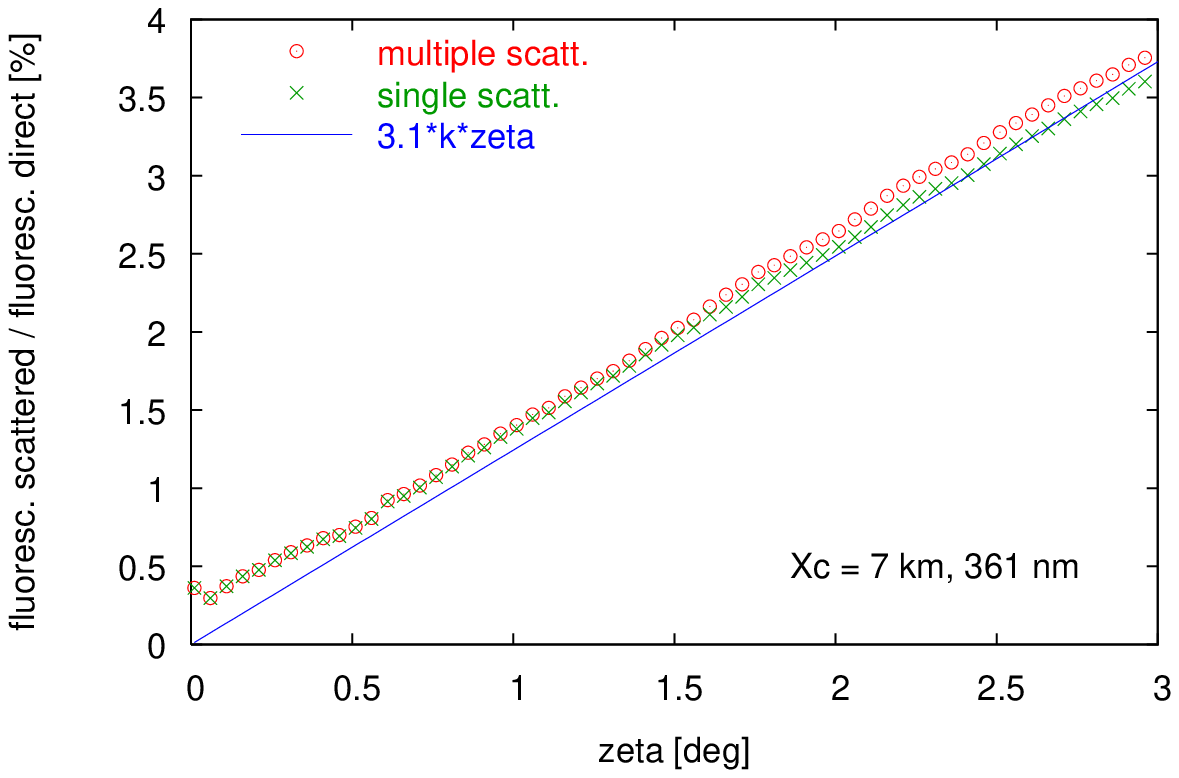}\\
\includegraphics[scale=0.95]{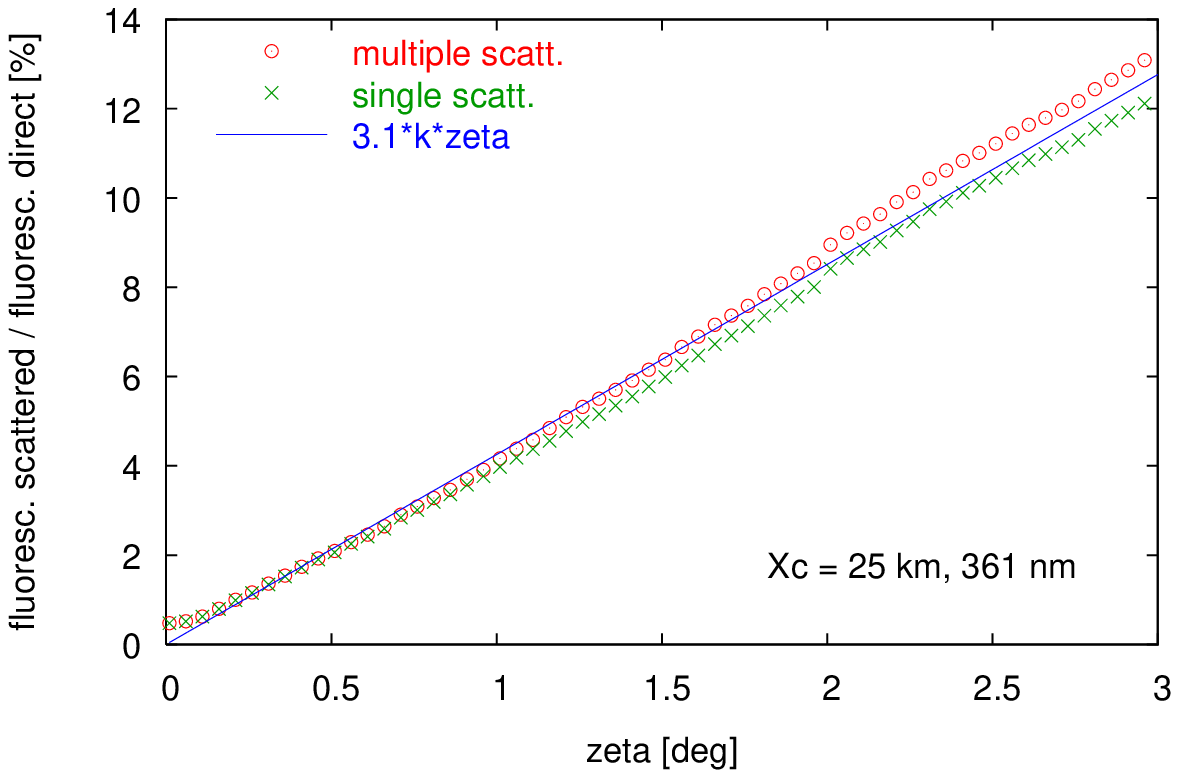}\\
\caption{\label {fig12} Comparison of results from our simulations of Rayleigh scattering only with the analytical function $3.1*k*\zeta$ (where $k=d/\lambda$) from \cite{giller}, for different shower-detector distances and wavelength bins.}
\end{center}
\end{figure}

Another comparison was made of our simulations with analytical calculations of Ref. \cite{giller}. These analytical calculations are done for Rayleigh scattered fluorescence light, assuming a constant-density atmosphere. To make the comparison, our simulation program was modified - allowed was Rayleigh scattering only. To simulate the effect of the constant-density atmosphere, for comparison were chosen points low above the horizon ($2^{\circ}$). However, one must remember that in a 3-dimensional simulation the photons traverse layers of atmosphere with different densities; also proximity to the ground level (elimination of photons at altitude zero) may influence the results. Although both calculations were done using completely different methods, final results are in a satisfying agreement (Fig. \ref{fig12}). The analytical formula of \cite{giller} for correction is also consistent with our fitted function - the multiple scattering contribution scales linearly with $\zeta$ angle and distance.

\section{Conclusions}

A Monte Carlo method for simulation of the multiple scattering of light has been developed, and used to determine the effect of this new contribution on the observation of air showers. Simulated is the development of air showers in a realistic atmosphere. Calculated is the contribution of the \textit{new signal}: scattered (both singly and multiply) fluorescence photons and multiply scattered Cherenkov photons to the observed shower image. The simulation results show that the new contribution to the signal recorded by a detector scales linearly with the signal collecting angle $\zeta$ and the shower-detector distance (geometrical distance $d$ or optical depth $\tau$), and falls exponentially with the shower front altitude above the ground; no dependence on the shower age is seen.

Different distributions of molecular atmosphere and aerosols, or angular distributions of light emission by an air shower were used in the simulations. It can be concluded that the size of the new signal depends on the distance between the shower and the detector, on the altitude of the observed point above ground, on the signal collecting angle in the detector and on the amount of aerosols in the air. It has been shown that one doesn't need to account separately for the variations of the molecular atmosphere profile, the vertical scale height of aerosol distribution, Cherenkov angular distribution or different altitudes (above sea level) of the detector.

As a result of this study, two parameterizations of the multiple scattering contribution to shower signal have been obtained. These parameterizations are simple functions of parameters characterizing the location of the air shower and the atmospheric conditions. One parameterization (Eq. \ref{fit} with parameters given in Eq. \ref{param1}) is a function of the signal collecting angle $\zeta$, the shower altitude above ground $h$, the shower-detector distance $d$, and the horizontal attenuation length $\Lambda_{T}$ (at 361 nm). The other parameterization (Eq. \ref{fit2} with parameters given in Eq. \ref{param2}) allows one to calculate the \textit{new signal} based on $\zeta$, $h$ and the optical depth along the shower-detector line of sight in the atmosphere $\tau$. These parameterizations can be easily implemented into existing reconstruction procedures used in analyses of air shower observations.

Scattering of fluorescence light and multiple scattering of Cherenkov photons makes a contribution to the signal received by the fluorescence detector, that should be taken into account in analysis of experimental data. This contribution can reach and exceed $\sim$ 10\% of the shower signal. Since the multiple scattering contribution changes with altitude, applying a correction for this effect will slightly change the shape of shower profile (and with it the depth of shower maximum).

In the form as it is presented in this work, the scattering correction should be included into reconstruction procedures after geometry of air shower and signal received at the detector are calculated. Another correction for the lateral distribution of light in shower image \cite{gora} is also applied at this stage. (This correction accounts for the part of shower signal that is lost outside the signal collecting angle $\zeta$.) We may notice that, while these two corrections have opposite effects on reconstruction results (taking multiple scattering into account reduces the signal from a shower, whereas lateral distribution correction increases it), they only rarely cancel each other: the effect of multiple scattering is larger for the distant showers than for the nearby ones, which is opposite to the lateral distribution correction. The application of both corrections is important, as they decrease the systematical uncertainty of final results.

It should be considered that in future not only shower reconstructions, but also simulations should include the effect of atmospheric multiple scattering. Also, in order to get consistent results, a number of calibration procedures may have to be amended to account for multiple scattering. Some methods of detector calibration and measurements of atmospheric conditions are based on observations of a distant light source or laser beam. Results of such measurements can also be influenced by the multiple scattering.

\textit{Acknowledgements.} We would like to thank Ralph Engel and Michael Unger for useful discussions and help in preparing this work, and Dariusz G\'ora and Markus Risse for collaboration at earlier stages of this work. This work was partially supported by the Polish Ministry of Science and Higher Education under grants N202 090 31/0623 and PAP/218/2006 and in Germany by the DAAD under grant No. 323-PPP.

\end{document}